\documentclass[journal]{IEEEtran}
\usepackage{amsmath,graphicx}
\usepackage{makecell}
\usepackage{multirow} 
\usepackage{booktabs}  
\usepackage{threeparttable} 
\usepackage{color}
\usepackage{xcolor}
\usepackage{amsfonts}
\usepackage{enumitem}
\usepackage[justification=centering]{subcaption}

\usepackage{pifont}
\newcommand{\True}{\ding{52}}
\newcommand{\False}{\ding{56}}
\usepackage{amsthm}

\usepackage[super]{nth}
\usepackage{engord} 
\usepackage{hyperref}

\newcommand{\myref}[1]{Eq.~\eqref{#1}}

\ifCLASSINFOpdf
\else
\fi
\usepackage{url}

\begin{document}
	%
	
	\title{ Single Image Deraining with Continuous Rain Density Estimation }
	%
	%
	%
	
	\author{Jingwei~He, Lei~Yu,~\IEEEmembership{Member,~IEEE}, Gui-Song~Xia,~\IEEEmembership{Senior Member,~IEEE} and Wen~Yang,~\IEEEmembership{Senior Member,~IEEE}
		\thanks{The research was partially supported by the National Natural Science
			Foundation of China under Grants 61871297.}
		\thanks{Jingwei~He, Lei~Yu and Wen~Yang are with
			School of Electronic Information, Wuhan University, Wuhan 430072, China
			(email: jingwei\_he@whu.edu.cn, ly.wd@whu.edu.cn,
			yangwen@whu.edu.cn).}
		\thanks{Gui-Song~Xia is with School of Computer Science, Wuhan University, Wuhan 430072, China
			(email: guisong.xia@whu.edu.cn).}
	}

	\maketitle
	\begin{abstract}
		Single image deraining (SIDR) often suffers from over/under deraining due to the nonuniformity of rain densities and the variety of raindrop scales. In this paper, we propose a \textbf{\it co}ntinuous \textbf{\it de}nsity guided network (CODE-Net) for SIDR. Particularly, it is composed of { a rain {\color{black}streak} extractor and a denoiser}, where the convolutional sparse coding (CSC) is exploited to filter out noises from the extracted rain streaks. Inspired by  the reweighted iterative soft-threshold for CSC, we address the problem of continuous rain density estimation by learning the weights with channel attention blocks from sparse codes. We further {\color{black}develop} a multiscale strategy to depict rain streaks appearing at different scales. Experiments on synthetic and real-world data demonstrate the superiority of our methods over recent {\color{black}state of the arts}, in terms of both quantitative and qualitative results. Additionally, instead of quantizing rain density with several levels, our CODE-Net can provide continuous-valued estimations of rain densities, which is more desirable in real applications.
	\end{abstract}
	
	\begin{IEEEkeywords}
		Image deraining, rain density estimation, convolutional sparse coding.
	\end{IEEEkeywords}

	%
	\IEEEpeerreviewmaketitle

	\section{Introduction}
	
	\label{sec:intro}
\begin{figure}[!t] 
	\centering
	\begin{subfigure}[t]{0.32\linewidth}
		\includegraphics[width=1.\linewidth]{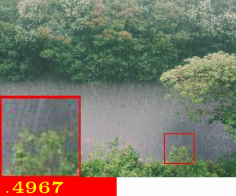}
		\includegraphics[width=1.\linewidth]{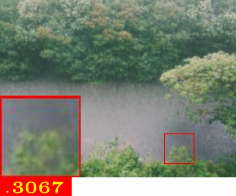}
		\includegraphics[width=1.\linewidth]{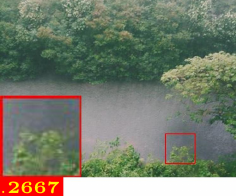}
		\subcaption{ light }
		\label{fig:real-light}
	\end{subfigure}
	\begin{subfigure}[t]{0.32\linewidth}
		\includegraphics[width=1.\linewidth]{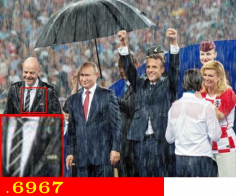}
		\includegraphics[width=1.\linewidth]{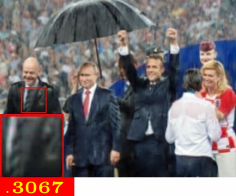}
		\includegraphics[width=1.\linewidth]{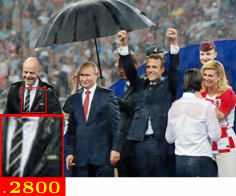}
		\subcaption{medium}
		\label{fig:real-medium}
	\end{subfigure}
	\begin{subfigure}[t]{0.32\linewidth}
		\includegraphics[width=1.\linewidth]{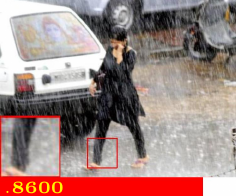}
		\includegraphics[width=1.\linewidth]{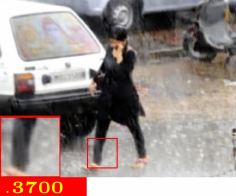}
		\includegraphics[width=1.\linewidth]{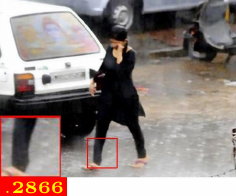}
		\subcaption{heavy  }
		\label{fig:real-heavy}
	\end{subfigure} 
	\caption{Deraining results on real images with light, medium and heavy rain (top) by SPANet \cite{spanet} (middle) and by our method (bottom).  {Our method could produce more pleasant results. The rain density of each image is estimated by our method and indicated in read.}
	 }
	\label{fig:1-real}
	\vspace{-0.2cm}
\end{figure}

	\IEEEPARstart{I}{mages} {captured in rainy weather are often severely contaminated by rain streaks, which may largely degrade the performance of the outdoor vision systems. Image deraining is to remove rain streaks from rainy images and many approaches have been proposed either {by adopting priors \cite{dsc,gmm,tmm2020lin} or by learning the inverse mapping from rainy to clean images \cite{cnn,ddn,jorder}.} However, the nonuniformity of rain densities and the variety of raindrop scales are still challenging problems and have rarely been considered even for the state-of-the-arts.  As shown in Fig.~\ref{fig:1-real}, for real images with light, medium and heavy rain densities and different raindrop scales, most of the state-of-the-art methods often suffer from over deraining (Fig.~\ref{fig:real-light} and \ref{fig:real-medium}) or under deraining (Fig.~\ref{fig:real-heavy}) results.} 
	
	One strategy to avoid the above problems is to build a large dataset with various rain densities, such as \cite{ddn} and \cite{did-mdn}. Correspondingly, a complex network with sufficient learning capacity needs to be meticulously designed. And the realistic rainy images are of a wide variety and difficult to collect. Another strategy is to  train multiple models for different rain densities. However, this strategy lacks flexibility in practice as additional approach to determine which model is required, and when the rain intensity varies widely,  models would occupy lots of storage space. To avoid these issues, Zhang \textit{et al.} \cite{did-mdn} utilized a CNN classifier to determine the density level (light, medium or heavy) and used it to  guide a multi-stream dense network. However, it is not appropriate since the real-world rain density is continuous, and the discrete-continuous gap may also lead to under or over deraining results. Hence, the adaptive method to estimate continuous densities is scarce.

	{
		
		In this paper, we propose a continuous density guided network (CODE-Net) for single image deraining (SIDR), which is capable to estimate the continuous rain densities, as shown in Fig.~\ref{fig:1-real}. Particularly, we consider the deraining task as a blind separation of clean image and rain streaks. To achieve this end, we propose to extract rain streaks with a simple CNN extractor followed by a denoiser to remove artifacts, and then subtract the denoised rain streaks from the rainy image. For the denoiser, the {convolutional sparse coding (CSC) \cite{zeiler2010deconvolutional}} is exploited to depict rain steaks. Intrinsically, the sparse coefficients are related to the rain density where the rain streaks with light density exhibits a simpler structure and thus results in sparser coefficients than with heavy density. Based on above inspirations,  we propose to take into account the rain density in two aspects:
		
		\begin{itemize}[noitemsep]
			\item We propose to learn the rain densities from the learned sparse coefficients through {channel attention (CA) \cite{rcan,senet} blocks}.
			\item By adaptively weighting the sparse coefficients,  the rain density can be implicitly considered. To achieve this end, {the weighted $\ell_1$-minimization \cite{reweighted}} is exploited with the weights learned according to rain densities, \textit{i.e.} sparse coefficients.
		\end{itemize}
		
		By imposing weighted sparsity, the denoiser can adaptively leverage the penalty for different rain densities. For the rain with light density, CSC produces sparser coefficients and thus leads to larger weights than with heavy density. Then with larger weights, the weighted $\ell_1$-minimization penalizes more on the coefficients and thus thresholds more disturbances. As a result, the denoiser is able to achieve better performance than without considering rain densities and thus give better deraining result, as shown in Fig.~\ref{fig:1-real}.

		It is completely different from \cite{did-mdn} where a CNN classifier is exploited to determine light, medium or heavy rain, we investigate  the relation between the rain density and the sparsity of rain streaks under the learned dictionaries. The CODE-Net can estimate the continuous rain densities. As shown in Fig.~\ref{fig:1-real}, the rain densities of different rain images are given with continuous value. Additionally, it even has potentials to measure the rain density of a given image to evaluate the deraining algorithms, as shown in Fig.~\ref{fig:1-real}.
		
		On the other hand, considering the variety of raindrop scales, we further exploit a multiscale strategy to depict rain steaks appearing at different scales and propose a multiscale CODE-Net (mCODE-Net).
		
	}

	In summary, our contributions are as follows:
	\begin{itemize}[noitemsep]
		\item We propose a density guided network (CODE-Net) and its multi-scale version mCODE-Net for single image deraining, where the rain density and raindrop scales are both considered. To the best of the authors' knowledge, it is the first work that can simultaneously tackle both problems for SIDR in one network.
		\item We propose an adaptive estimator for the continuous valued rain density evaluation and various experiments are implemented to validate its effectiveness. Thus the proposed CODE-Net can be exploited to quantitatively evaluate the performance of deraining for real rainy images.   
		\item Experiments on synthetic and real-world data demonstrate the superiority of the proposed methods over recent state-of-the-arts, in terms of both quantitative and qualitative results. And meanwhile, some possible extensions with CODE-Net to deraining with perceptual loss and to other image enhancing tasks, \textit{i.e.} desnowing are given.  
		
	\end{itemize}
	
	The remainder of this paper is organized as follows. We first introduce the related work in Section  \ref{background}. Then, a detailed description of our methodology is presented in Section \ref{sec:promethod}. In Section \ref{sec:exper}, various experimental results demonstrate the effectiveness of our method. Finally, we conclude the paper and discuss the future work in Section \ref{sec:Conclusion}.
	
	\section{Related Work} \label{background}
	In this section, we present a brief review on recent single image de-raining (SIDR) methods, including prior-based ones and deep CNN models.
	
	\textbf{Prior-based methods.} In the past decades, to make the inverse problem SIDR well-posed, various priors on natural clean background or rain streaks are employed. Thanks to the boosting of Compressed Sensing \cite{donoho2006compressed}, Kang \textit{et al.} \cite{kang2011automatic} and Luo \textit{et al.} \cite{dsc} proposed to separate the rain streaks via {sparse prior}. Observing that rain streaks typically span a narrow range of directions, Zhu \textit{et al.} \cite{zhu2017joint} introduced a
	rain removal method based on the local gradient statistics, in addition to sparse prior. In \cite{chen2013generalized}, Chen \textit{et al.} presented a low-rank representation-based framework for rain streak removal. Besides, there are several methods based on patch-based GMM priors \cite{gmm} and nonlocal self-similarity priors \cite{kim2013single}. However, the above methods tend to produce over-smoothed results \cite{kang2011automatic,zhu2017joint,gmm}. What's more, the performance of them depends heavily on the hand-crafted priors, which means they would struggle to cope with real images with complicated scenes.

	\textbf{Deep CNN models.} More recently, the progress was attributed to deep CNN models, which directly learned the nonlinear mapping between rainy and clean images in a data-driven manner. Fu \textit{et al.} \cite{cnn} firstly decomposed a rainy image into a low-pass base layer and a high-pass detail layer, and then performed a CNN-based deraining operation on the detail layer. Built on \cite{cnn},  Fu \textit{et al.} extended the network to its residual version in \cite{ddn}. In \cite{jorder}, Yang \textit{et al.}  introduced a multi-task network for joint rain detection and removal. Zhang \textit{et al.} \cite{did-mdn} created a density-aware multi-stream densely connected convolutional neural network for SIDR. To generalize to real scenes, Wei \textit{et al.} \cite{semi} presented a semi-supervised learning paradigm that combined synthetic and real images in training stage. Wang \textit{et al.} \cite{spanet} proposed a spatial attentive model based on direction-aware attention mechanism. Besides, the recurrent learning  was also {employed to consider} the dependencies of deep features across stages in \cite{prenet} and \cite{rescan}.
	
	\begin{figure*}[!t]
		\centering
		\includegraphics[width=0.99\linewidth]{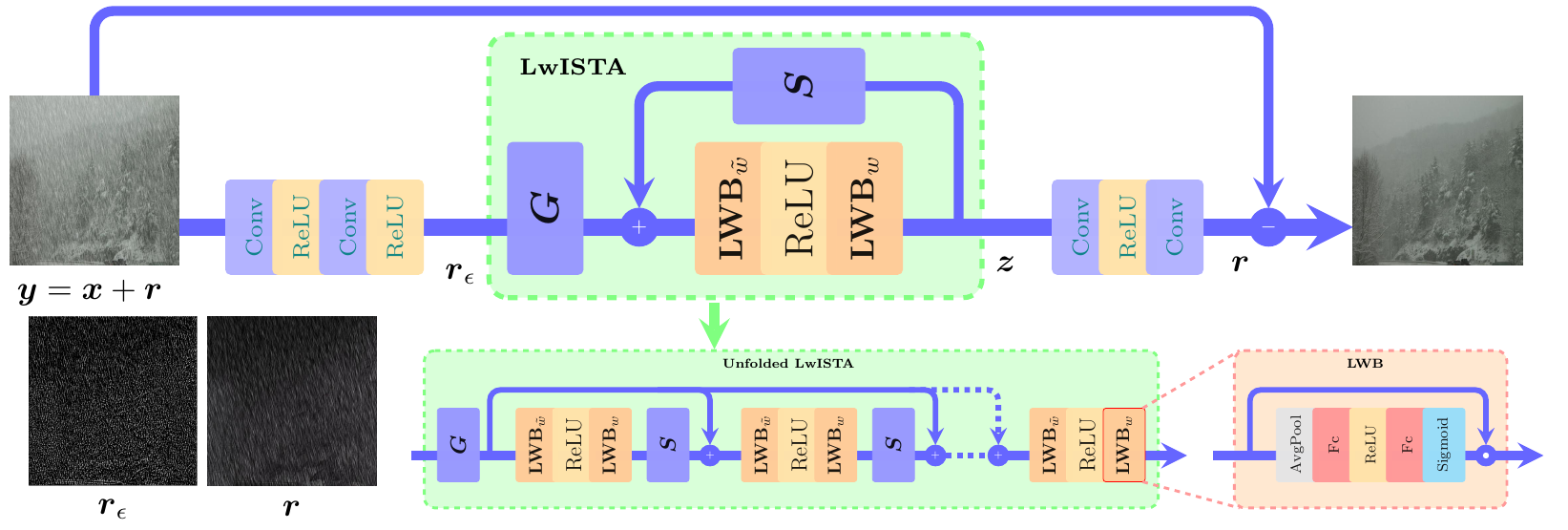}
		\caption{Architecture of the proposed CODE-Net and its sub-modules, LwISTA and LWB.}
		\label{fig:model}
	\end{figure*}

	{Most existing algorithms ignored the nonuniformity of rain densities and the variety of raindrop scales, {\color{black}except \cite{did-mdn} which classifies rain into light, medium or heavy levels and leverages a multi-stream
	dense network to guide the rain removal.} However, such simple classification may be problematic in reality since the real-world rain state is continuous, which is verified in Section \ref{derain-comparison}. Different from these methods, we propose to learn the rain density from the sparse coefficients under the framework of convolutional sparse coding (CSC) \cite{zeiler2010deconvolutional}. Besides, our network can be easily extended to {\color{black}a multiscale version} to consider rain densities and scales simultaneously.}

	\section{Image Deraining via CODE-Net}
	\label{sec:promethod}
	{\color{black}Mathematically, the rainy image $ \boldsymbol{y} \in \mathbb{R}^{m \times n} $ can be modeled as a linear superimposition of the clean background $ \boldsymbol{x} \in \mathbb{R}^{m \times n} $ and the rain streaks component $ \boldsymbol{r} \in \mathbb{R}^{m \times n}$:
		\begin{equation}\label{eq:1}
		\centering
		\boldsymbol{y} =  \boldsymbol{x}+\boldsymbol{r},
		\end{equation}
		and the goal of SIDR is to reconstruct $ \boldsymbol{x} $ from $ \boldsymbol{y} $. According to the rainy imaging model \myref{eq:1}, one can find that an accurate reconstruction of rain streaks $\boldsymbol{r}$ plays an important role for image deraining. Moreover, rain streaks are often with regular structures and thus exploiting the classical convolutional sparse  coding (CSC).}
	
	\subsection{Weighted CSC Model for Rain Streaks}
	\label{denoising_by_lwista}
	{\color{black}Under the framework of CSC}, one can express the mathematical model for rain streaks:
	\begin{equation}\label{eq:2}
	\boldsymbol{r}_\epsilon =\sum_{i=1}^{c} \boldsymbol{f}_{i} \otimes \boldsymbol{z}_{i} + \boldsymbol{\epsilon}  
	\end{equation}
	{where $\left\{\boldsymbol{z}_{i}\right\}_{i=1}^{c} \in \mathbb{R}^{c \times l_h \times l_w} $  are $ c $ {sparse coefficients} of noisy rain streak map $ \boldsymbol{r}_\epsilon \in \mathbb{R}^{1\times l_h \times l_w} $ \textit{w.r.t.} corresponding rain kernels $\left\{\boldsymbol{f}_{i}\right\}_{i=1}^{c} \in \mathbb{R}^{c \times s \times s} $, $\boldsymbol{\epsilon}$ denotes the noise and $ \otimes $ denotes the convolution operation. }
	
	To yield $ \boldsymbol{z}_{i} $, a common solution is to solve the following $\ell_{1}$-minimization problem:
	\begin{equation} \label{equ:l1csc}
	\arg \min _{\boldsymbol{z}} \frac{1}{2}\left\|\boldsymbol{r}_\epsilon-\sum_{i=1}^{c} \boldsymbol{f}_{i} \otimes \boldsymbol{z}_{i}\right\|_{2}^{2}+\lambda \sum_{i=1}^{c} \left\|\boldsymbol{z}_{i}\right\|_{1},
	\end{equation}
	where $  \|\cdot\|_{p} $ is $\ell_{p}$-norm and $\lambda$ is a regularization parameter balancing sparsity and fidelity. 
	
	{Intrinsically, the sparse coefficients could reflect the rain density since the light rain exhibits a simpler structure and thus results in coefficients sparser than heavy rain. To take into account the rain density implicitly, we introduce a weight parameter $ w_i $ for $ \boldsymbol{z}_i $ \cite{reweighted}}:
	\begin{equation} \label{equ:rwcsc}
	\arg \min _{\boldsymbol{z}} \frac{1}{2}\left\|\boldsymbol{r}_\epsilon-\sum_{i=1}^{c} \boldsymbol{f}_{i} \otimes \boldsymbol{z}_{i}\right\|_{2}^{2}+\lambda \sum_{i=1}^{c} {w}_{i}\left\|\boldsymbol{z}_{i}\right\|_{1}.
	\end{equation}
	
	Due to the  fact that convolution operation can be replaced with matrix multiplication, \myref{equ:rwcsc} can be reformulated as:
	\begin{equation}\label{equ:reformulated-rwcsc}
	\arg \min _{ \boldsymbol{z}} \frac{1}{2}\left\|\boldsymbol{r}_\epsilon-\boldsymbol{F}\boldsymbol{z}\right\|_{2}^{2}+\lambda \sum_{i=1}^{c}w_{i}\left\|\boldsymbol{z}_{i}\right\|_{1},
	\end{equation}
	where $ \boldsymbol{F}=[\boldsymbol{F}_{1},\boldsymbol{F}_{2},\dots,\boldsymbol{F}_{c}] $ {\color{black}is the concatenation of convolution matrices corresponding to filters $\boldsymbol{f}_{i}$},  $ \boldsymbol{z}=[\boldsymbol{z}_{1},\boldsymbol{z}_{2},\dots,\boldsymbol{z}_{c}] $, and $ \boldsymbol{F}_{i}\boldsymbol{z}_{i} \equiv \boldsymbol{f}_{i} \otimes \boldsymbol{z}_{i} $.  Note that $ \boldsymbol{r}_\epsilon $ and $ \boldsymbol{z}_{i}$ are ordered lexicographically as column vectors. From this view, CSC can be regarded as a special case of conventional sparse coding. {\color{black}Thus the solution to \myref{equ:reformulated-rwcsc} can be obtained by following iterations \cite{reweighted}}:
	\begin{equation}  \label{equ:rwista}
	\begin{split}
	\boldsymbol{z}^{(t+1)}&=\Gamma_{\boldsymbol{w}^{(t)}\frac{\lambda}{L}}\left(\boldsymbol{z}^{(t)}+\frac{1}{L} \boldsymbol{F}^{T}\left(\boldsymbol{r}_\epsilon-\boldsymbol{F} \boldsymbol{z}^{(t)}\right)\right)\\
	&=\Gamma_{\boldsymbol{w}^{(t)}\frac{\lambda}{L}}\left(\boldsymbol{S} 
	\otimes	\boldsymbol{z}^{(t)} + \boldsymbol{G} \otimes \boldsymbol{r}_\epsilon\right),
	\end{split}
	\end{equation}
	where $ \boldsymbol{w} = [w_{1},w_{2},\dots,w_{c}] \in \left(0,1 \right] $, $ \Gamma_{\theta}(\cdot) $ denotes the element-wise soft thresholding function, defined by $ \Gamma_{\theta}(\alpha) = \text{sign}(\alpha)\text{max}(|\alpha|-\theta,0) $, $\boldsymbol{I} $ is the identity matrix, and $L$ is the Lipschitz constant. Note that, since $\boldsymbol{I} $ and  $\boldsymbol{F} $ are  sparse convolution matrices, there must be $ \boldsymbol{S} $ and $ \boldsymbol{G} $ satisfying $ \boldsymbol{S} \otimes	\boldsymbol{z}^{(t)}=\left( \boldsymbol{I}-\frac{1}{L} \boldsymbol{F}^{T}\boldsymbol{F}\right)\boldsymbol{z}^{(t)}$ and $ \boldsymbol{G} \otimes	\boldsymbol{r}_\epsilon= \frac{1}{L} \boldsymbol{F}^{T}\boldsymbol{r}_\epsilon$.
	
	{
		{ Since $ \Gamma_{\theta}(\cdot) $ is an element-wise operation and $ \boldsymbol{w} > 0 $, $ \Gamma_{\theta}(\cdot) $ has an important attribute:}
		\begin{equation} \label{attribute_of_gamma_function}
		\begin{split}
		\Gamma_{\boldsymbol{w} \theta}\left( \boldsymbol{\alpha}  \right)
		&=\text{sign}(\boldsymbol{\alpha} )\text{max}(\left| \boldsymbol{\alpha} \right|-\boldsymbol{w} \theta,0)\\
		&=\boldsymbol{w}\odot \text{sign}(\boldsymbol{\alpha} )\text{max}(\left| \boldsymbol{\tilde{w}}\odot \boldsymbol{\alpha} \right|- \theta,0)\\
		&=\boldsymbol{w}\odot \text{sign}(\boldsymbol{\tilde{w}}\odot\boldsymbol{\alpha} )\text{max}(\left| \boldsymbol{\tilde{w}}\odot \boldsymbol{\alpha} \right|- \theta,0)\\
		&=\boldsymbol{w}\odot \Gamma_{ \theta}\left(\boldsymbol{\tilde{w}}\odot \boldsymbol{\alpha}  \right),
		\end{split}
		\end{equation}
		where $ \boldsymbol{\tilde{w}}=[\frac{1}{{w}_{1}},\frac{1}{{w}_{2}},\dots,\frac{1}{{w}_{c}}] $, and $\odot$ denotes the element-wise product.
		Based on such observation, we rewrite \myref{equ:rwista} as:
		\begin{equation} \label{rewrite_final_solution_to_reweight_csc}
		\begin{split}
		\boldsymbol{v}^{(t)} &=\boldsymbol{S} \otimes \boldsymbol{z}^{(t)} + \boldsymbol{G} \otimes \boldsymbol{r}_\epsilon   \\
		\boldsymbol{z}^{(t+1)}
		&=\boldsymbol{w}^{(t)} \odot \Gamma_{\theta}\left(\boldsymbol{\tilde{w}}^{(t)} \odot \boldsymbol{v}^{(t)}\right).
		\end{split}
		\end{equation}
	}

	In \myref{rewrite_final_solution_to_reweight_csc}, $ \boldsymbol{\tilde{w}}  $ is mainly used to adjust sparsity of $\boldsymbol{z}$ while $\boldsymbol{w}$ for scaling. Large $ \boldsymbol{\tilde{w}}  $ favors less sparse solution, and vice versa. {Namely, $ \boldsymbol{\tilde{w}}  $ is conducive to more accurate representations of rain with various densities, resulting in better deraining results. {On the other hand}, $ \boldsymbol{\tilde{w}}  $ implies  the rain density and thus could be utilized to estimate rain density.}

	\subsection{{\color{black}Continuous Density Guided Deraining by Weights}}
	\label{rde-with-weights}
	Recall the CSC model for rain streaks \myref{eq:2}, the basic idea behind \myref{eq:2} is to sparsely represent the rain streaks with learned kernels $\{\boldsymbol{f}_i\}_{i=1}^c$. Intrinsically, the light rain is supposed to have more concise representation than the heavy rain due to the fact of simpler appearance. To this end, we choose the channel attention (CA) as our learning weight block (LWB), defined by:
	\begin{equation} \label{def:lwb}
	\text{LWB} \left( \boldsymbol{\alpha} \right)  =\boldsymbol{\alpha} \odot  \underbrace{\delta \left( {Fc}_{2} \left(  \text{ReLU}  \left(  {Fc}_{1}  \left( \text{AvgPool}\left( {\boldsymbol{\alpha}}\right) \right)\right) \right) \right)}_{\text{weight}},
	\end{equation}
	where $ Fc_{1} $ and $  Fc_{2} $ are full connection layers without biases, ReLU denotes the Rectified Linear Unit activation function, $ \text{AvgPool}(\cdot):\mathbb{R}^{c \times  h \times w}\rightarrow \mathbb{R}^{c \times  1 \times 1}$ denotes 
	the global average pooling function, and $ \delta(\cdot) $ denotes the sigmoid function. 
	
	{Benefitting from \myref{def:lwb},  for light rain, small valued weights $\tilde{\boldsymbol{w}}$ in \myref{rewrite_final_solution_to_reweight_csc} {\color{black}would be generated}, which means
	\myref{def:lwb} will penalize the sparse coefficients $\boldsymbol{z}$ {\color{black}much more} to reduce sparsity. On the contrary, for the heavy rain with less sparse representation under the learned kernels $\{\boldsymbol{f}_i\}_{i=1}^c$, the corresponding weights $\tilde{\boldsymbol{w}}$ will be large to enhance sparsity. As a consequence, {\color{black}by adaptively weighting the sparse coefficients in \myref{rewrite_final_solution_to_reweight_csc} using LWB \myref{def:lwb}}, the rain density can be implicitly considered, which proved in Section \ref{sec:exper} contributes to better rain removal results.}

\begin{figure}[!t] 
	\centering
	\begin{subfigure}[t]{1\linewidth}
		\begin{minipage}[!t]{1\linewidth}
			\begin{subfigure}[b]{0.3\linewidth}
				\includegraphics[width=1.\linewidth,height=1\linewidth]{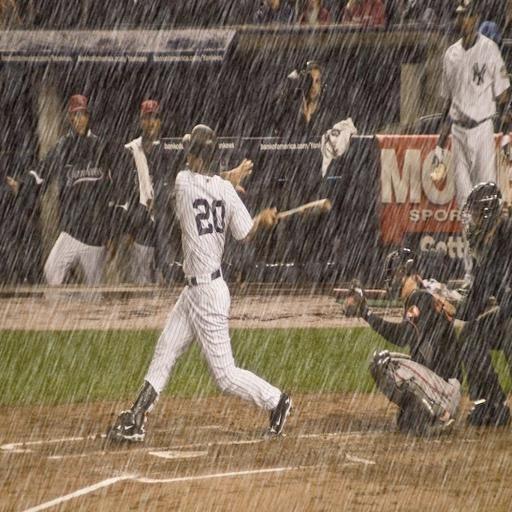}
				\includegraphics[width=1.\linewidth,trim=5 0 20 0,clip]{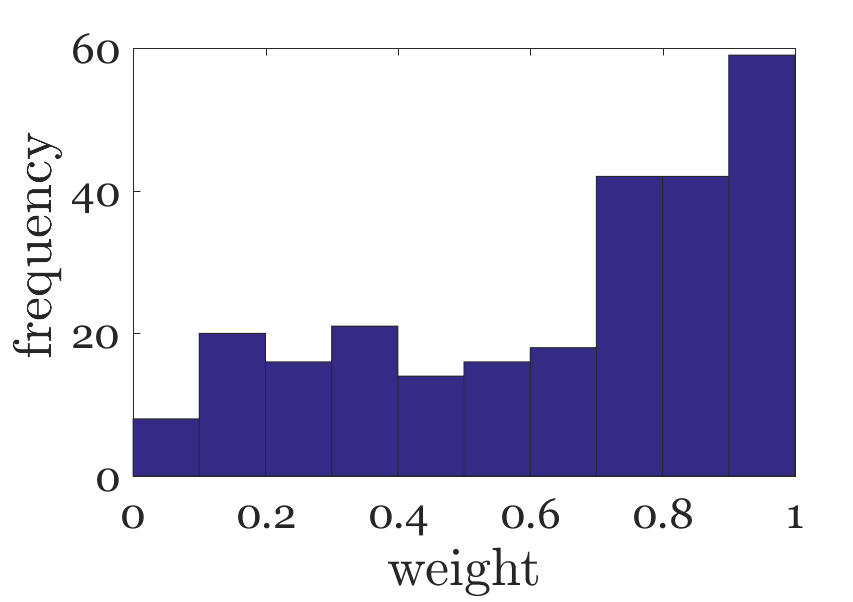}
				\subcaption*{ heavy}
			\end{subfigure}
			\begin{subfigure}[b]{0.3\linewidth}
				\includegraphics[width=1.\linewidth,height=1\linewidth]{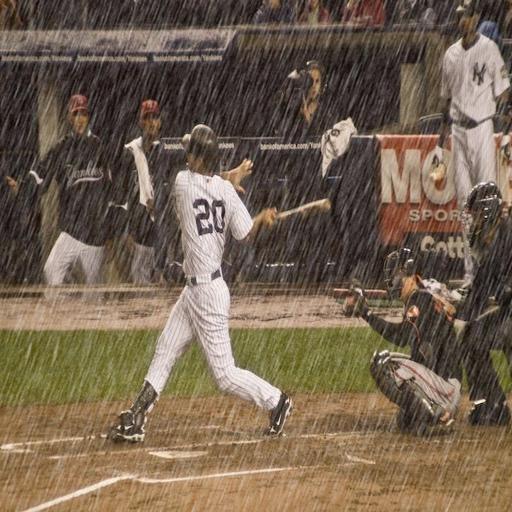}
				\includegraphics[width=1.\linewidth,trim=5 0 20 0,clip]{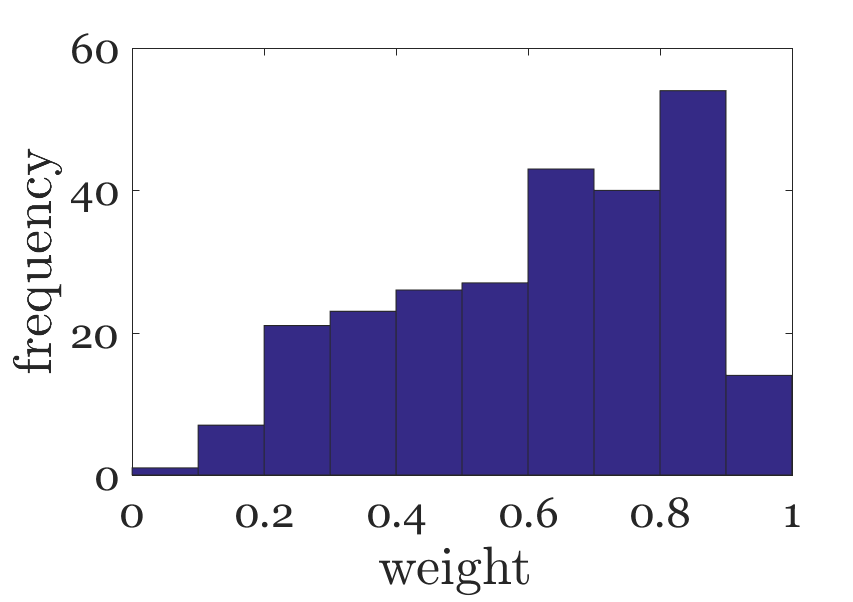}
				\subcaption*{ medium}
			\end{subfigure}
			\begin{subfigure}[b]{0.3\linewidth}
				\includegraphics[width=1.\linewidth,height=1\linewidth]{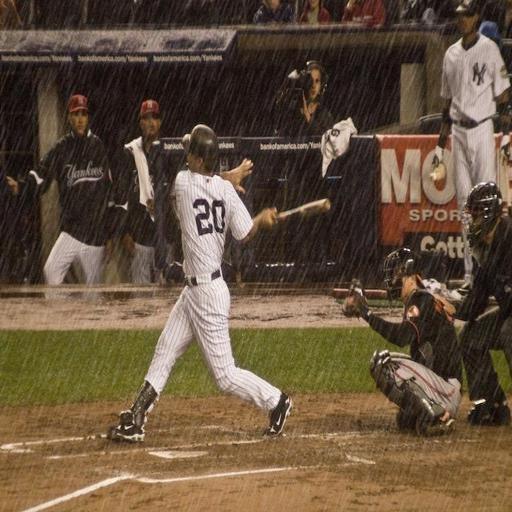}
				\includegraphics[width=1.\linewidth,trim=5 0 20 0,clip]{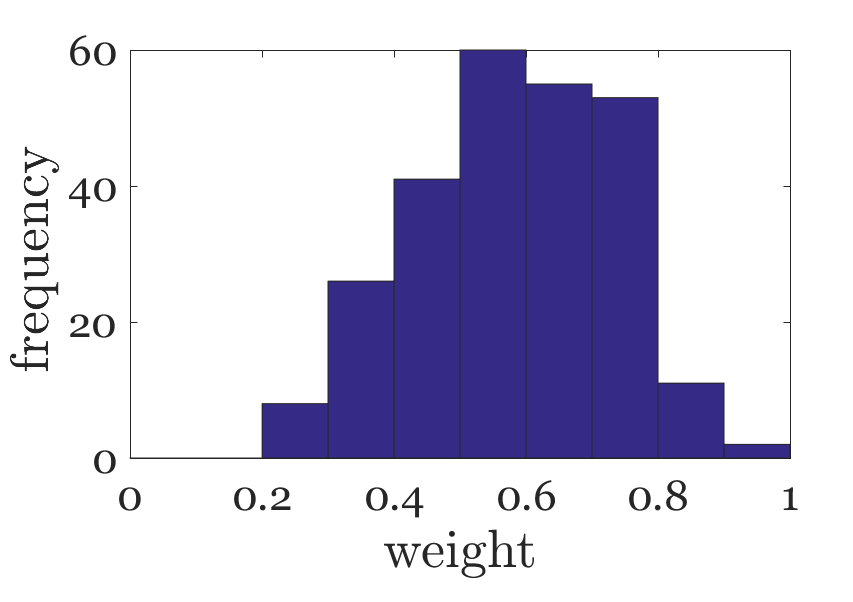}
				\subcaption*{ light}
			\end{subfigure}
		\end{minipage}
		\subcaption{ {a sample from \cite{did-mdn} with heavy, medium and light rain densities, and the corresponding distribution of learned weights $\tilde{\boldsymbol{w}}$ in \myref{rewrite_final_solution_to_reweight_csc}.} }
		\label{fig:example-rain1200}
	\end{subfigure}\\
	\centering
	\begin{subfigure}[t]{1\linewidth}
\centering
		\includegraphics[width=0.7\linewidth]{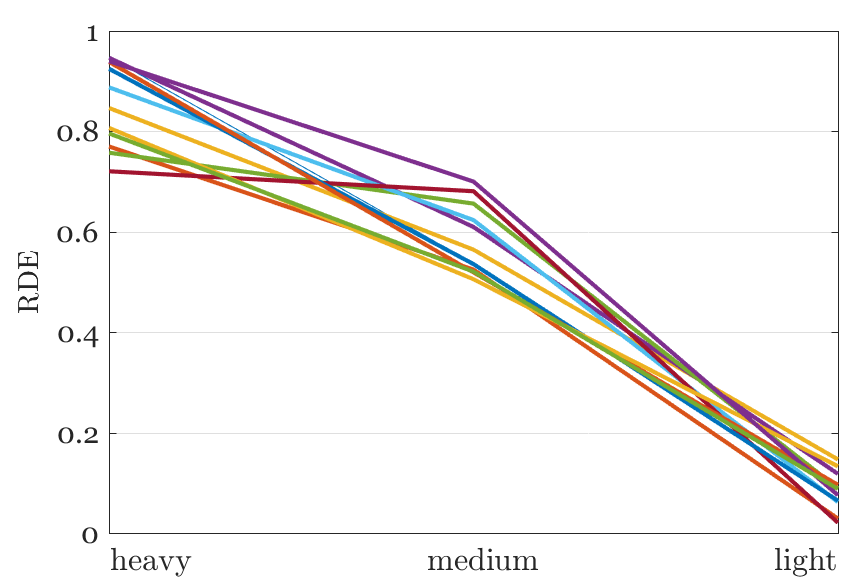}
		\subcaption{ RDEs of different samples (colored lines).}
		\label{fig:analysis-weight}
	\end{subfigure}
	\caption{ A sample (a) and several RDEs (b) of different rain levels from Rain1200 \cite{did-mdn}.}
	\label{fig:examples-and-average-weights}
	
\end{figure}

	{
		 { {On the other hand}, since the rain density is closely related to the weights $\tilde{\boldsymbol{w}}$, we could estimate the rain density from $\tilde{\boldsymbol{w}}$.}
		As shown in Fig.~\ref{fig:example-rain1200}, we illustrate the distribution of learned weights $\tilde{\boldsymbol{w}}$ in \myref{rewrite_final_solution_to_reweight_csc}, one can see that the weight cluster gradually approaches 1 with the increase of rain intensity. Therefore, we define the average $\tilde{\boldsymbol{w}}$ of different channels as rain density estimation (RDE):
		\begin{equation} \label{rain-density-estimation}
		\text{RDE}  = \dfrac{1}{c} \sum_{i=1}^{c} \tilde{\boldsymbol{w}}_i,
		\end{equation}
		the more it rains, the greater RDE should be, as shown in Fig.~\ref{fig:analysis-weight}. And RDEs of real raining images could be found in Section \ref{subsec:rain-density-estimation}. {\color{black}Note that all RDEs are normalized between zero and one.}
		
		Note that the weight is updated adaptively according to \myref{def:lwb}. It implies that the rain density is estimated  without any labels in our model, instead of training an extra network in a supervised way to classify the rain density into three categories in DID-MDN \cite{did-mdn}. Besides, thanks to the weights,  our model is capable of estimating the rain density with continuous states, which is more suitable for real raining scenes than algorithms only classifying rain density into limited discrete states, for instance, DID-MDN \cite{did-mdn}. 
	}

\begin{figure*}[!t] 
	\includegraphics[width=0.99\linewidth]{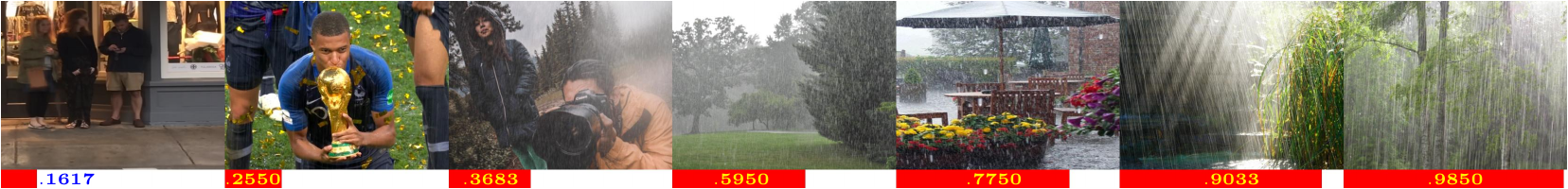}
	\caption{ A clear image ({{blue}}) and several samples ({{yellow}}) of different rain levels from  \cite{spanet} and corresponding RDEs.}
	\label{fig:real-rie-all}
\end{figure*}
	\begin{figure*}[!t] 
	\centering
	\begin{subfigure}[t]{0.2\linewidth}
		\includegraphics[width=1.\linewidth]{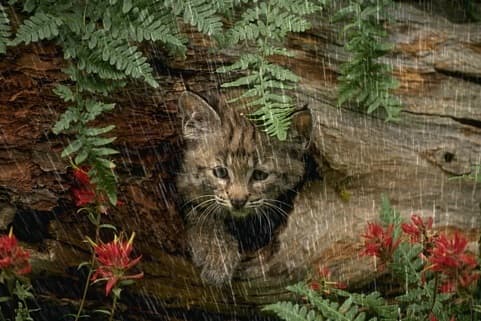}
		\vspace*{-0.5cm}
		\subcaption*{Input \\ (25.34, 0.8402)  }
	\end{subfigure}
	\hspace{-0.112cm}
	\begin{subfigure}[t]{0.2\linewidth}
		\includegraphics[width=1.\linewidth]{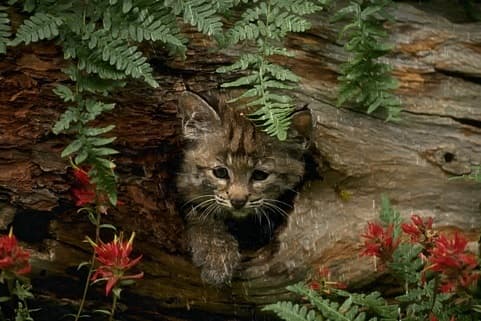}
		\vspace*{-0.5cm}
		\subcaption*{DDN \cite{ddn} \\ (30.51, 0.9147)  }
	\end{subfigure}
	\hspace{-0.112cm}
	\begin{subfigure}[t]{0.2\linewidth}
		\includegraphics[width=1.\linewidth]{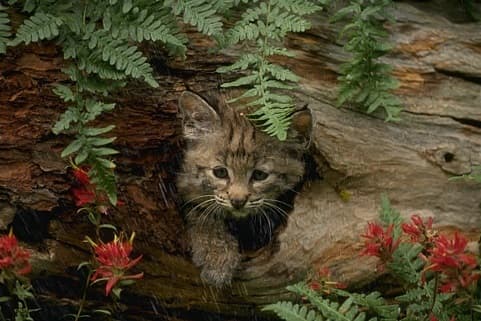}
		\vspace*{-0.5cm}
		\subcaption*{JORDER \cite{jorder} \\  (\underline{32.37}, 0.9206)}
	\end{subfigure}
	\hspace{-0.112cm}
	\begin{subfigure}[t]{0.2\linewidth}
		\includegraphics[width=1.\linewidth]{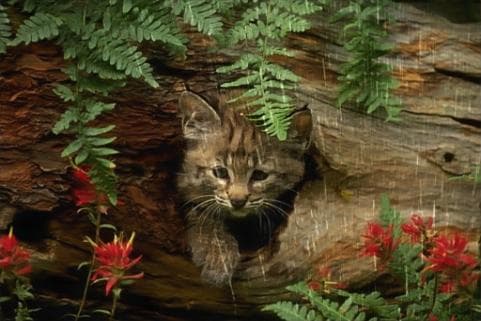}
		\vspace*{-0.5cm}
		\subcaption*{  DID-MDN \cite{did-mdn} \\ (29.31, 0.8537) }
	\end{subfigure}
	\hspace{-0.112cm}
	\begin{subfigure}[t]{0.2\linewidth}
		\includegraphics[width=1.\linewidth]{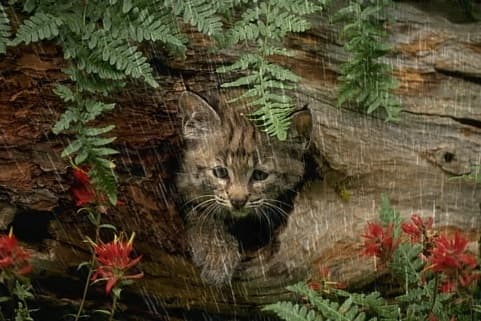}
		\vspace*{-0.5cm}
		\subcaption*{SPANet \cite{spanet} \\ (28.49, 0.8663) }
	\end{subfigure}
	\hspace{-0.112cm}
	\begin{subfigure}[t]{0.2\linewidth}
		\includegraphics[width=1.\linewidth]{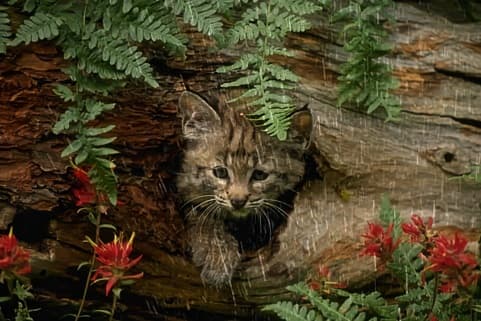}
		\vspace*{-0.5cm}
		\subcaption*{PreNet \cite{prenet} \\ (30.15, 0.8902) }
	\end{subfigure}
	\hspace{-0.112cm}
	\begin{subfigure}[t]{0.2\linewidth}
		\includegraphics[width=1.\linewidth]{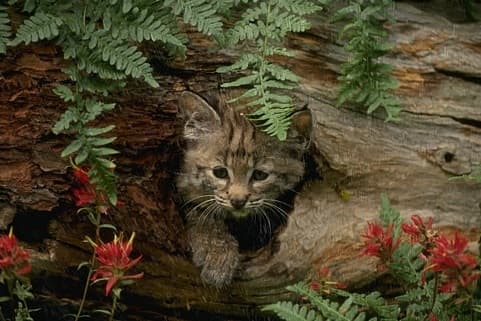}
		\vspace*{-0.5cm}
		\subcaption*{CODE-Net (ours) \\ ({32.25}, \underline{0.9216})  }
	\end{subfigure}
	\hspace{-0.112cm}
	\begin{subfigure}[t]{0.2\linewidth}
		\includegraphics[width=1.\linewidth]{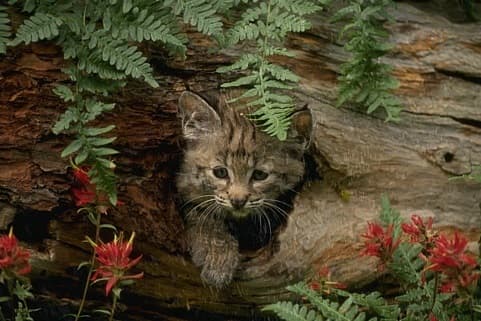}
		\vspace*{-0.5cm}
		\subcaption*{mCODE-Net (ours) \\ (\textbf{32.51}, \textbf{0.9317})  }
	\end{subfigure}\\
    \vspace*{0.1cm}
	\centering
	\begin{subfigure}[t]{0.2\linewidth}
		\includegraphics[width=1.\linewidth]{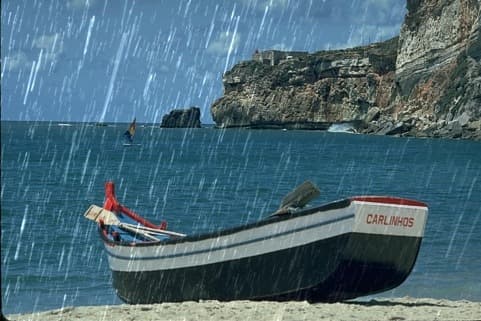}
		\vspace*{-0.5cm}
		\subcaption*{	Input \\ (27.55, 0.7977)}
	\end{subfigure}
	\hspace{-0.12cm}
		\begin{subfigure}[t]{0.2\linewidth}
		\includegraphics[width=1.\linewidth]{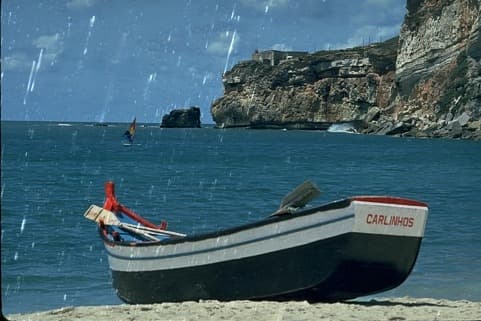}
		\vspace*{-0.5cm}
		\subcaption*{	DDN \cite{ddn} \\ (32.10, 0.9199)}
	\end{subfigure}
	\hspace{-0.12cm}
	\begin{subfigure}[t]{0.2\linewidth}
		\includegraphics[width=1.\linewidth]{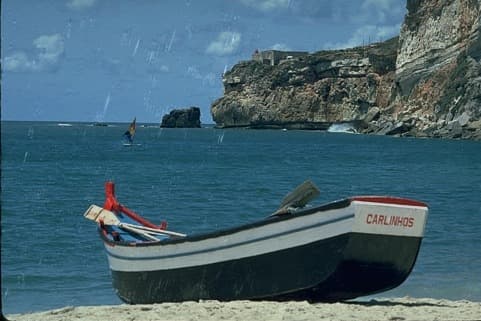}
		\vspace*{-0.5cm}
		\subcaption*{JORDER \cite{jorder} \\ (36.83, 0.9556)}
	\end{subfigure}
	\hspace{-0.12cm}
	\begin{subfigure}[t]{0.2\linewidth}
		\includegraphics[width=1.\linewidth]{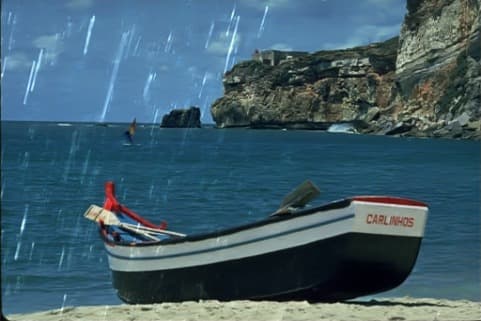}
		\vspace*{-0.5cm}
		\subcaption*{DID-MDN \cite{did-mdn} \\ (28.43, 0.8876)}
	\end{subfigure}
	\hspace{-0.12cm}
		\begin{subfigure}[t]{0.2\linewidth}
		\includegraphics[width=1.\linewidth]{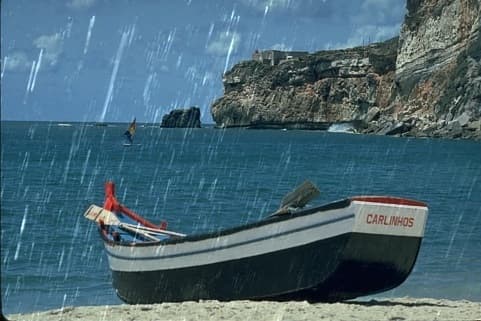}
		\vspace*{-0.5cm}
		\subcaption*{SPANet \cite{spanet} \\ (29.96, 0.8748)}
	\end{subfigure} 
	\hspace{-0.12cm}
	\begin{subfigure}[t]{0.2\linewidth}
		\includegraphics[width=1.\linewidth]{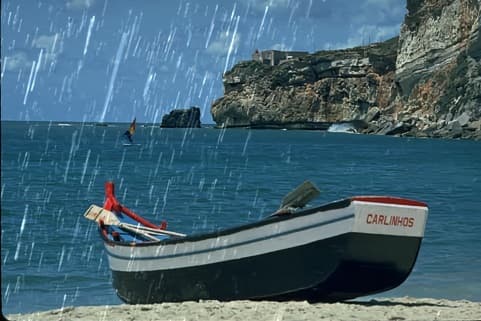}
		\vspace*{-0.5cm}
		\subcaption*{PreNet \cite{prenet} \\ (28.90, 0.8454)}
	\end{subfigure} 
	\hspace{-0.12cm}
	\begin{subfigure}[t]{0.2\linewidth}
		\includegraphics[width=1.\linewidth]{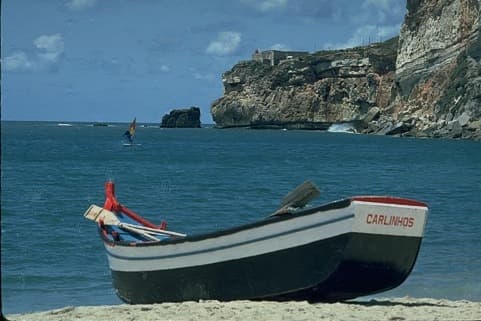}
		\vspace*{-0.5cm}
		\subcaption*{ CODE-Net (ours) \\ ({\underline{37.28}}, \textbf{0.9678})}
	\end{subfigure}
	\hspace{-0.12cm}
	\begin{subfigure}[t]{0.2\linewidth}
		\includegraphics[width=1.\linewidth]{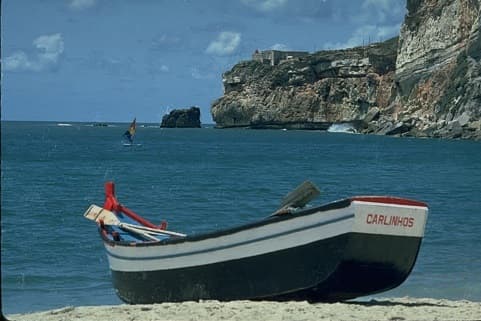}
		\vspace*{-0.5cm}
		\subcaption*{mCODE-Net (ours) \\ ( \textbf{37.67}, \underline{0.9670})}
	\end{subfigure} 
	
	\caption{Qualitative comparisons (PSNR, SSIM) of two synthetic images from Rain12.}
	\label{Qualitative_comparison_on_synthetic_rain12}
\end{figure*}

	\begin{figure*}[!t]
	\centering
	\begin{subfigure}[t]{0.2\linewidth}
		\includegraphics[width=1.\linewidth,height=0.8\linewidth]{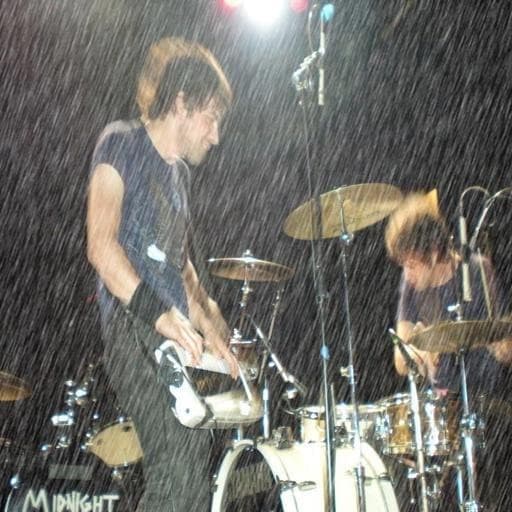}
		\vspace*{-0.5cm}
		\subcaption*{	Input \\ (20.59, 0.6078)}
	\end{subfigure}
	\hspace{-0.12cm}
	\begin{subfigure}[t]{0.2\linewidth}
		\includegraphics[width=1.\linewidth,height=0.8\linewidth]{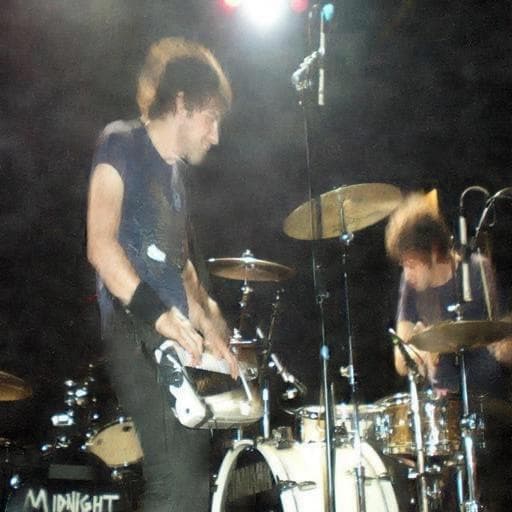}
		\vspace*{-0.5cm}
		\subcaption*{	DDN \cite{ddn} \\ (29.89, 0.8860)}
	\end{subfigure}
	\hspace{-0.12cm}
	\begin{subfigure}[t]{0.2\linewidth}
		\includegraphics[width=1.\linewidth,height=0.8\linewidth]{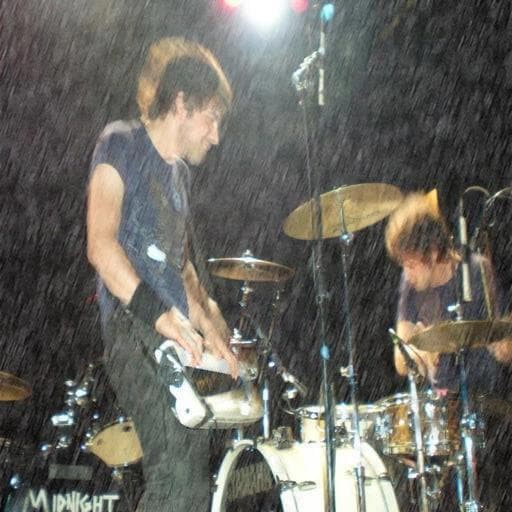}
		\vspace*{-0.5cm}
		\subcaption*{JORDER \cite{jorder} \\ (23.31, 0.7231)}
	\end{subfigure}
	\hspace{-0.12cm}
	\begin{subfigure}[t]{0.2\linewidth}
		\includegraphics[width=1.\linewidth,height=0.8\linewidth]{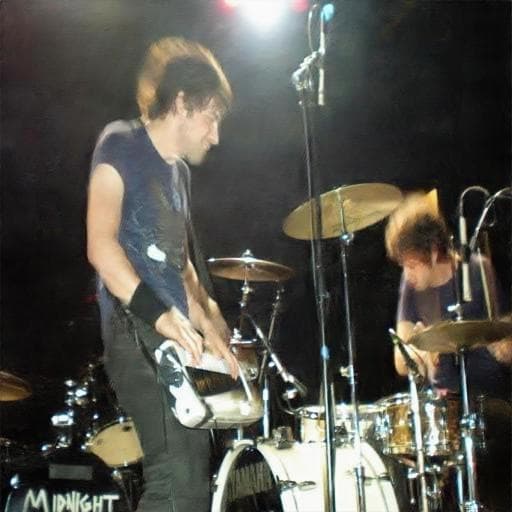}
		\vspace*{-0.5cm}
		\subcaption*{DID-MDN \cite{did-mdn} \\ (29.12, 0.9144)}
	\end{subfigure} 
	\hspace{-0.12cm}
		\begin{subfigure}[t]{0.2\linewidth}
		\includegraphics[width=1.\linewidth,height=0.8\linewidth]{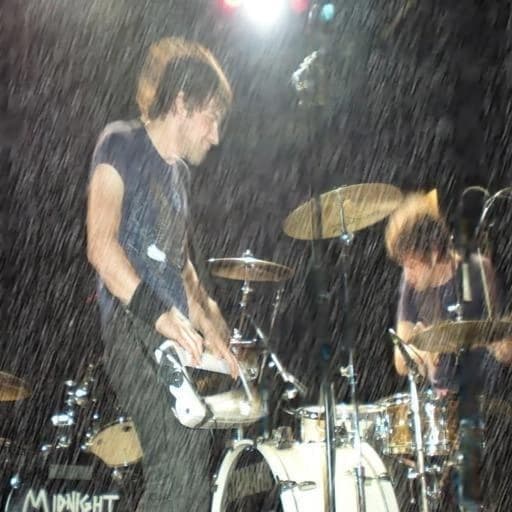}
		\vspace*{-0.5cm}
		\subcaption*{SPANet \cite{spanet} \\ (21.95, 0.6959)}
	\end{subfigure} 
	\hspace{-0.12cm}
	\begin{subfigure}[t]{0.2\linewidth}
		\includegraphics[width=1.\linewidth,height=0.8\linewidth]{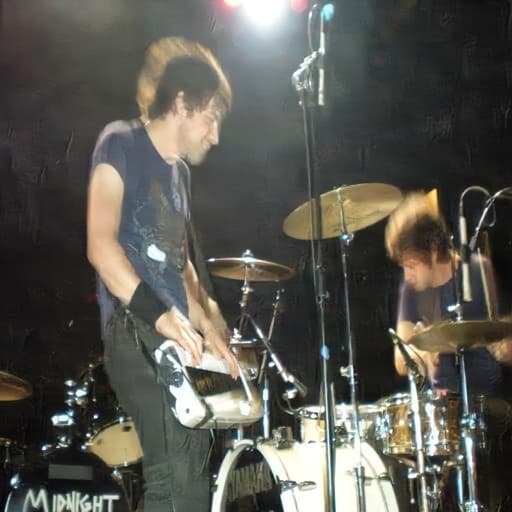}
		\vspace*{-0.5cm}
		\subcaption*{PreNet \cite{prenet} \\ (33.11, 0.9289)}
	\end{subfigure} 
	\hspace{-0.12cm}
	\begin{subfigure}[t]{0.2\linewidth}
		\includegraphics[width=1.\linewidth,height=0.8\linewidth]{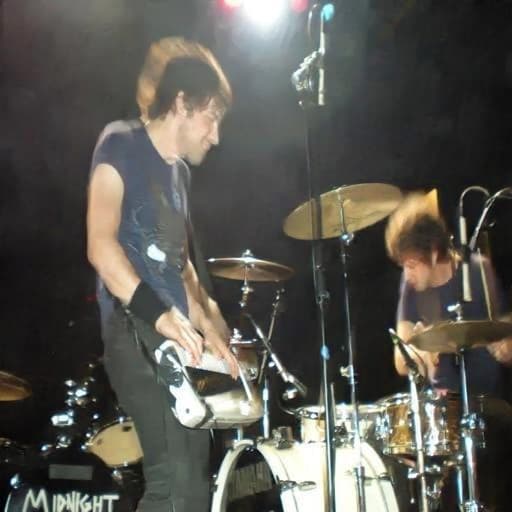}
		\vspace*{-0.5cm}
		\subcaption*{ CODE-Net (ours) \\ ({\underline{33.21}}, \underline{0.9328})}
	\end{subfigure}
	\hspace{-0.12cm}
	\begin{subfigure}[t]{0.2\linewidth}
		\includegraphics[width=1.\linewidth,height=0.8\linewidth]{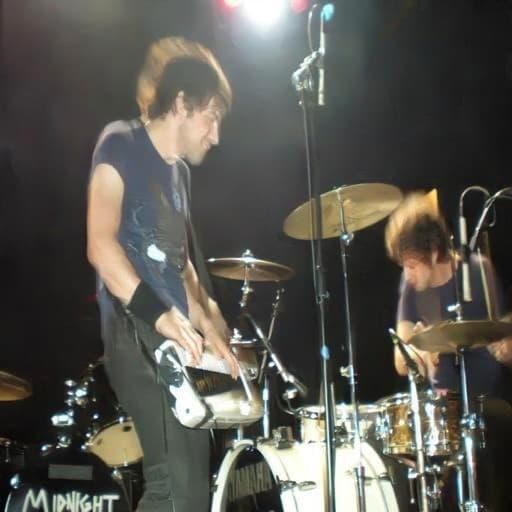}
		\vspace*{-0.5cm}
		\subcaption*{mCODE-Net (ours) \\ ( \textbf{34.29}, \textbf{0.9443})}
	\end{subfigure} \\
	\vspace*{0.1cm}
	\centering
	\begin{subfigure}[t]{0.2\linewidth}
		\includegraphics[width=1.\linewidth,height=0.8\linewidth]{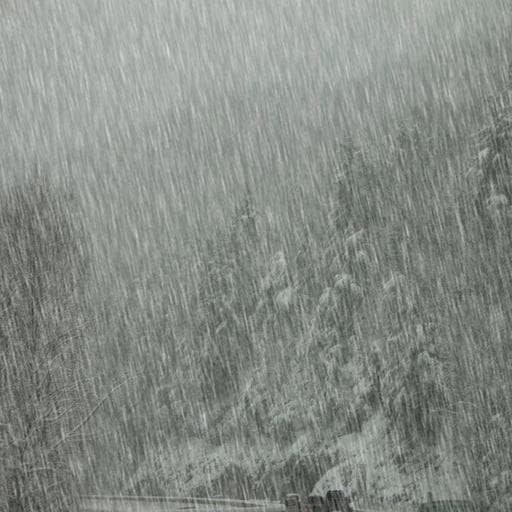}
		\vspace*{-0.5cm}
		\subcaption*{Input \\ (22.12, 0.6319)  }
	\end{subfigure}
	\hspace{-0.112cm}
	\begin{subfigure}[t]{0.2\linewidth}
		\includegraphics[width=1.\linewidth,height=0.8\linewidth]{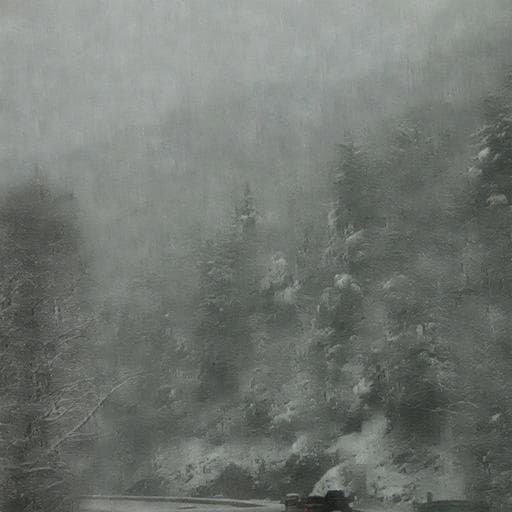}
		\vspace*{-0.5cm}
		\subcaption*{DDN \cite{ddn} \\ (31.21, 0.8465)  }
	\end{subfigure}
	\hspace{-0.112cm}
	\begin{subfigure}[t]{0.2\linewidth}
		\includegraphics[width=1.\linewidth,height=0.8\linewidth]{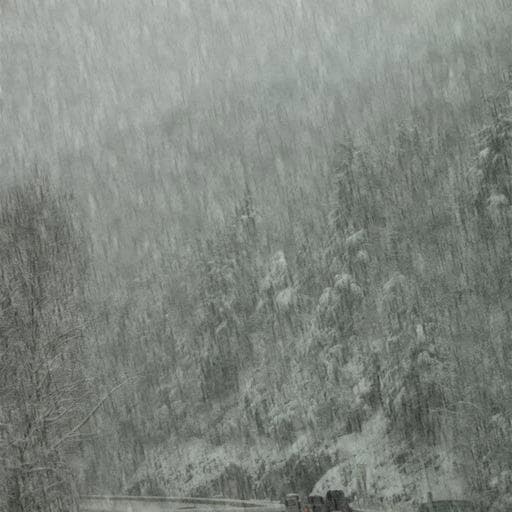}
		\vspace*{-0.5cm}
		\subcaption*{JORDER \cite{jorder} \\  (24.95, 0.7653)}
	\end{subfigure}
	\hspace{-0.112cm}
	\begin{subfigure}[t]{0.2\linewidth}
		\includegraphics[width=1.\linewidth,height=0.8\linewidth]{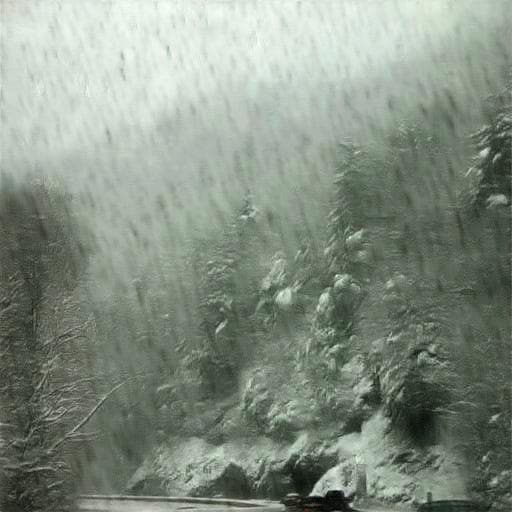}
		\vspace*{-0.5cm}
		\subcaption*{  DID-MDN \cite{did-mdn} \\ (18.25, 0.7647) }
	\end{subfigure}
	\hspace{-0.112cm}
		\begin{subfigure}[t]{0.2\linewidth}
		\includegraphics[width=1.\linewidth,height=0.8\linewidth]{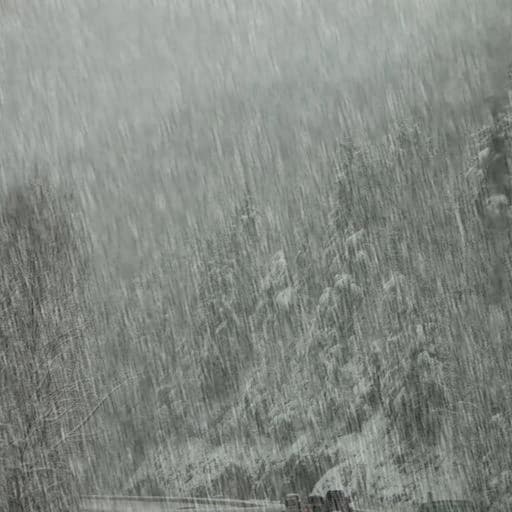}
		\vspace*{-0.5cm}
		\subcaption*{SPANet \cite{spanet} \\ (23.31, 0.7298) }
	\end{subfigure}
	\hspace{-0.112cm}
	\begin{subfigure}[t]{0.2\linewidth}
		\includegraphics[width=1.\linewidth,height=0.8\linewidth]{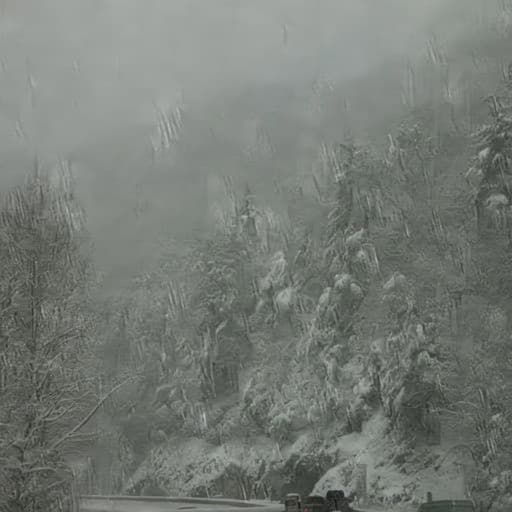}
		\vspace*{-0.5cm}
		\subcaption*{PreNet \cite{prenet} \\ (30.89, 0.8532) }
	\end{subfigure} 
	\hspace{-0.112cm}
	\begin{subfigure}[t]{0.2\linewidth}
		\includegraphics[width=1.\linewidth,height=0.8\linewidth]{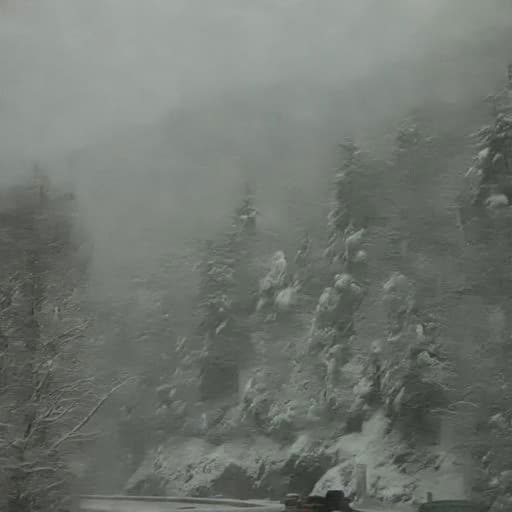}
		\vspace*{-0.5cm}
		\subcaption*{CODE-Net (ours) \\ (\underline{33.19}, \underline{0.8813})  }
	\end{subfigure}
	\hspace{-0.112cm}
	\begin{subfigure}[t]{0.2\linewidth}
		\includegraphics[width=1.\linewidth,height=0.8\linewidth]{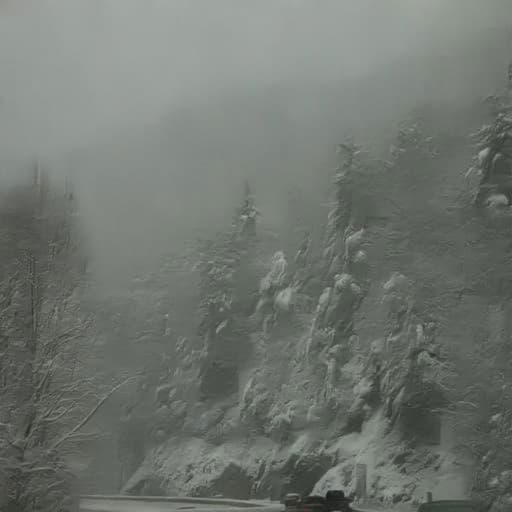}
		\vspace*{-0.5cm}
		\subcaption*{mCODE-Net (ours) \\ (\textbf{33.59}, \textbf{0.8918})  }
	\end{subfigure}
	
	\caption{Qualitative comparisons (PSNR, SSIM) of two synthetic images from Rain1200.}
	\label{Qualitative_comparison_on_synthetic_rain1200}
\end{figure*}

\begin{figure*}[!t]
	\centering
	\begin{subfigure}[t]{0.2\linewidth}
		\includegraphics[width=1.\linewidth,trim=0 127 0 0 ,clip]{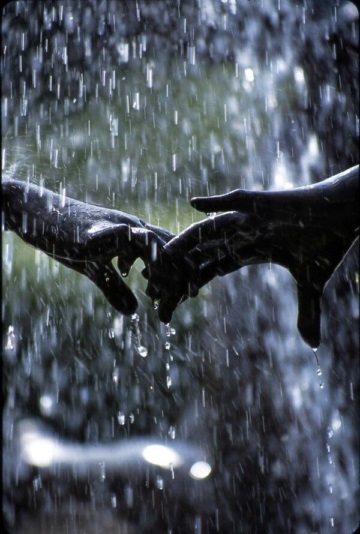} 
		\vspace*{-0.5cm}
		\subcaption*{Input \\0.4833}
	\end{subfigure}
	\hspace{-0.12cm}
	\begin{subfigure}[t]{0.2\linewidth}
		\includegraphics[width=1.\linewidth,trim=0 127 0 0 ,clip]{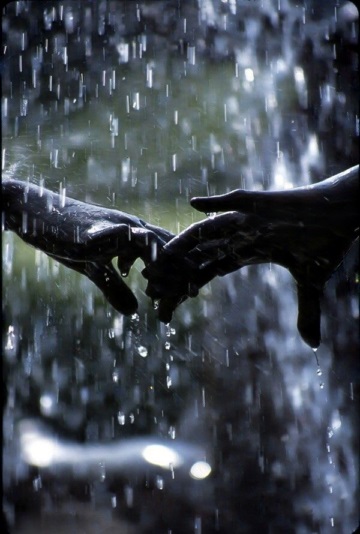} 
		\vspace*{-0.5cm}
		\subcaption*{DDN \cite{ddn} \\ 0.2433}
	\end{subfigure}
	\hspace{-0.12cm}
	\begin{subfigure}[t]{0.2\linewidth} 
		\includegraphics[width=1.\linewidth,trim=0 127 0 0 ,clip]{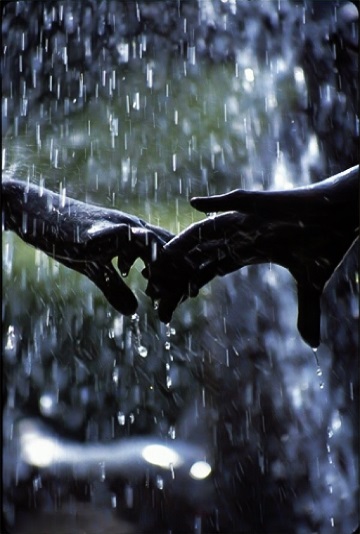}  
		\vspace*{-0.5cm}
		\subcaption*{DID-MDN \cite{did-mdn} \\ 0.3267}
	\end{subfigure} 
	\hspace{-0.12cm}
	\begin{subfigure}[t]{0.2\linewidth}
		\includegraphics[width=1.\linewidth,trim=0 127 0 0 ,clip]{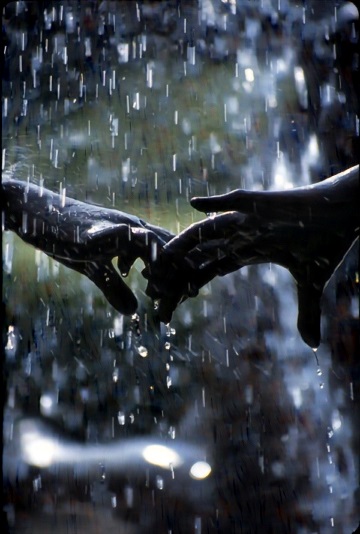}  
		\vspace*{-0.5cm}
		\subcaption*{SEMI \cite{semi} \\0.2467}
	\end{subfigure}
	\hspace{-0.12cm}
	\begin{subfigure}[t]{0.2\linewidth}
		\includegraphics[width=1.\linewidth]{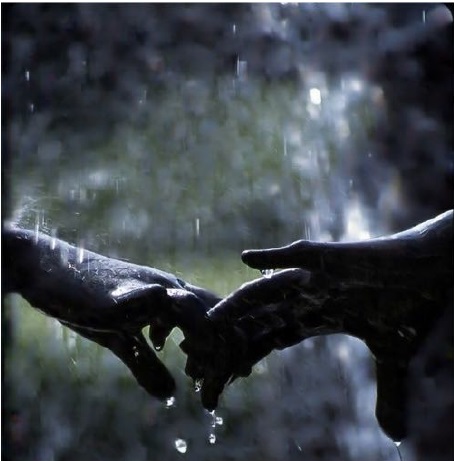}  
		\vspace*{-0.5cm}
		\subcaption*{SPANet \cite{spanet} \\0.1417}
	\end{subfigure}
	\hspace{-0.12cm}
	\begin{subfigure}[t]{0.2\linewidth}
		\includegraphics[width=1.\linewidth,trim=0 127 0 0 ,clip]{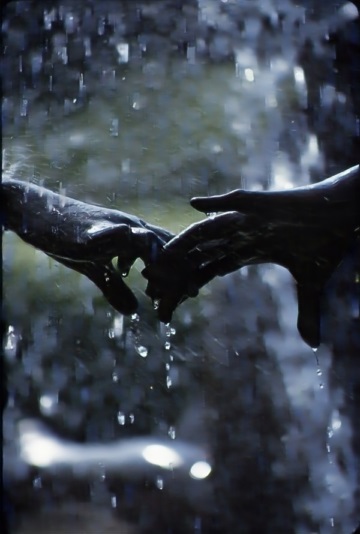}  
		\vspace*{-0.5cm}
		\subcaption*{PreNet \cite{prenet} \\0.1333}
	\end{subfigure}
	\hspace{-0.12cm}
	\begin{subfigure}[t]{0.2\linewidth}
		\includegraphics[width=1.\linewidth,trim=0 127 0 0 ,clip]{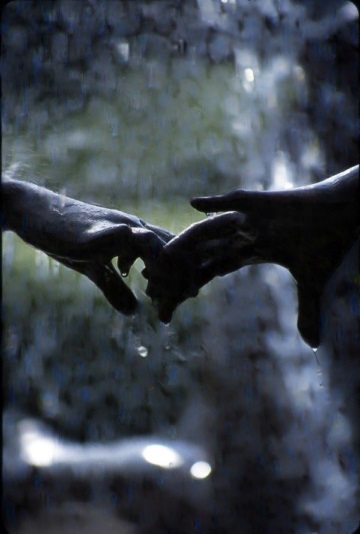}  
		\vspace*{-0.4cm}
		\subcaption*{CODE-Net (ours) \\0.1217}
	\end{subfigure}
	\hspace{-0.12cm}
	\begin{subfigure}[t]{0.2\linewidth}
		\includegraphics[width=1.\linewidth,trim=0 127 0 0 ,clip]{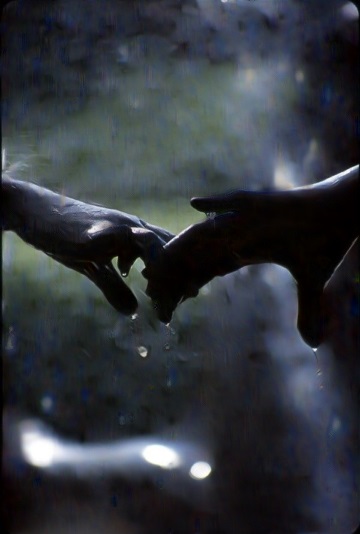}  
		\vspace*{-0.4cm}
		\subcaption*{mCODE-Net (ours) \\0.0667}
	\end{subfigure}\\	
	\centering
	\begin{subfigure}[t]{0.2\linewidth}
		\includegraphics[width=1.\linewidth]{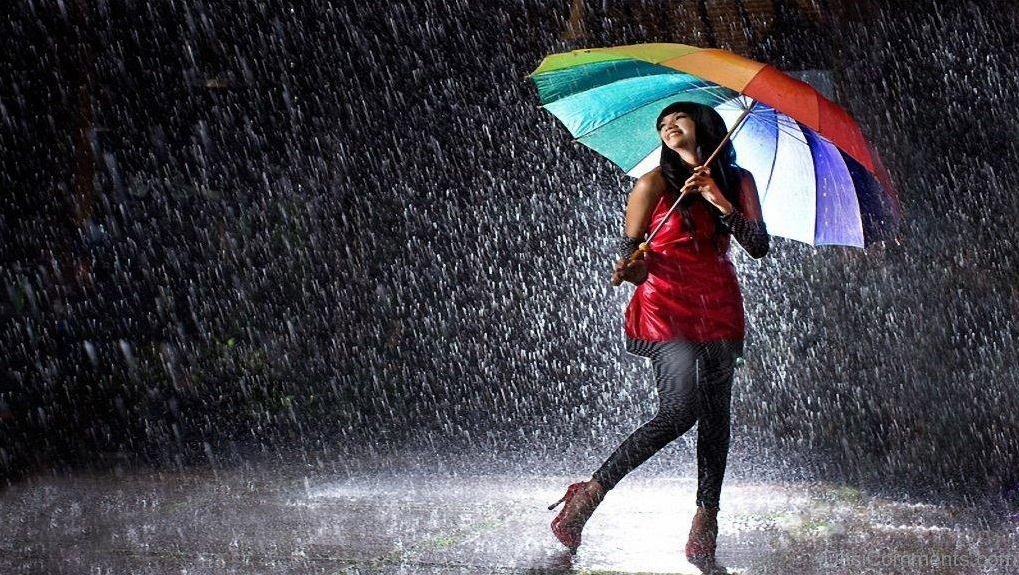} 
		\vspace*{-0.5cm}
		\subcaption*{Input \\ 0.5367}
	\end{subfigure}
	\hspace{-0.12cm}
	\begin{subfigure}[t]{0.2\linewidth}
		\includegraphics[width=1.\linewidth]{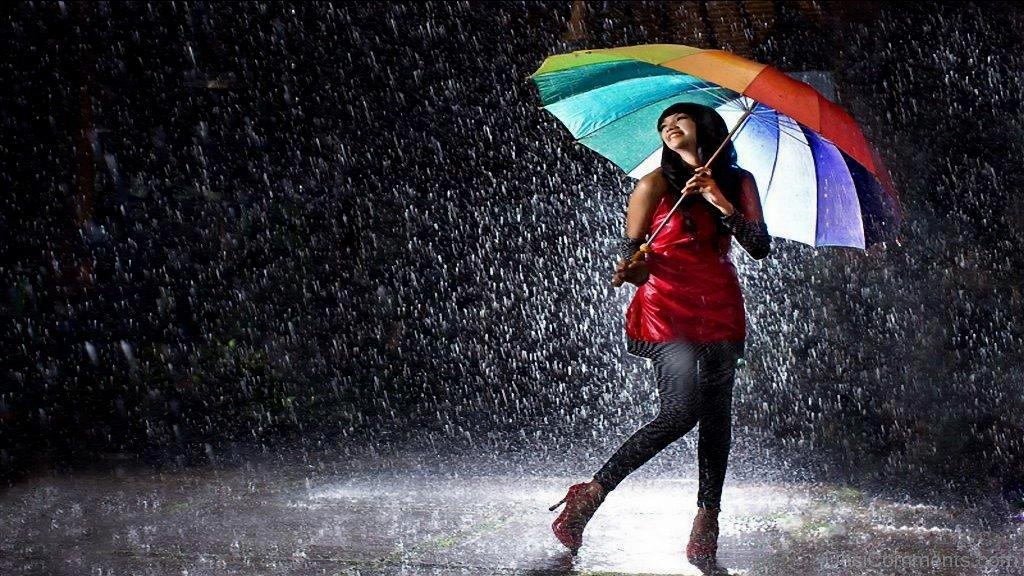} 
		\vspace*{-0.5cm}
		\subcaption*{DDN \cite{ddn} \\ 0.1967}
	\end{subfigure}
	\hspace{-0.12cm}
	\begin{subfigure}[t]{0.2\linewidth} 
		\includegraphics[width=1.\linewidth]{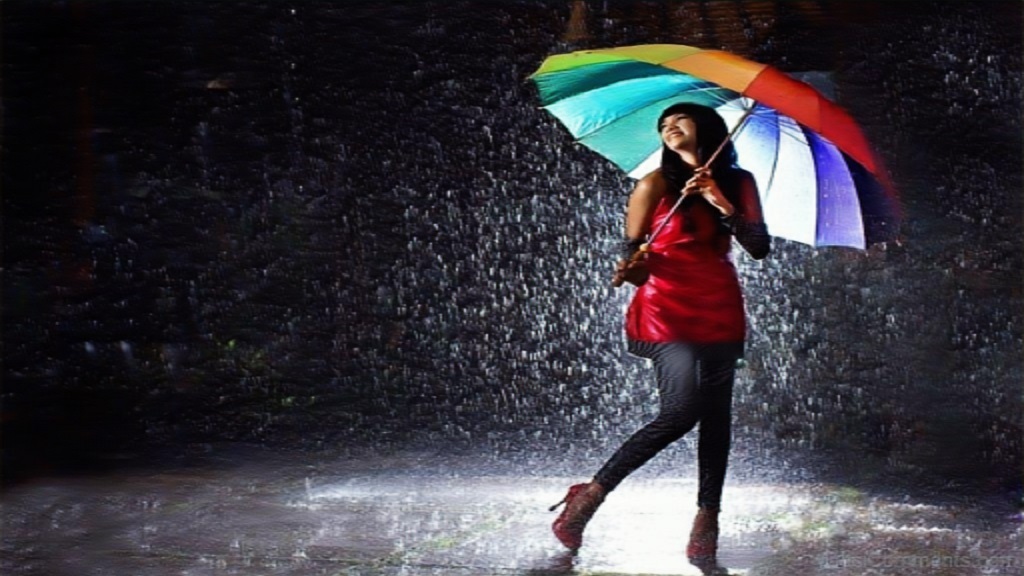}  
		\vspace*{-0.5cm}
		\subcaption*{DID-MDN \cite{did-mdn} \\ 0.1867}
	\end{subfigure} 
	\hspace{-0.12cm}
	\begin{subfigure}[t]{0.2\linewidth}
		\includegraphics[width=1.\linewidth]{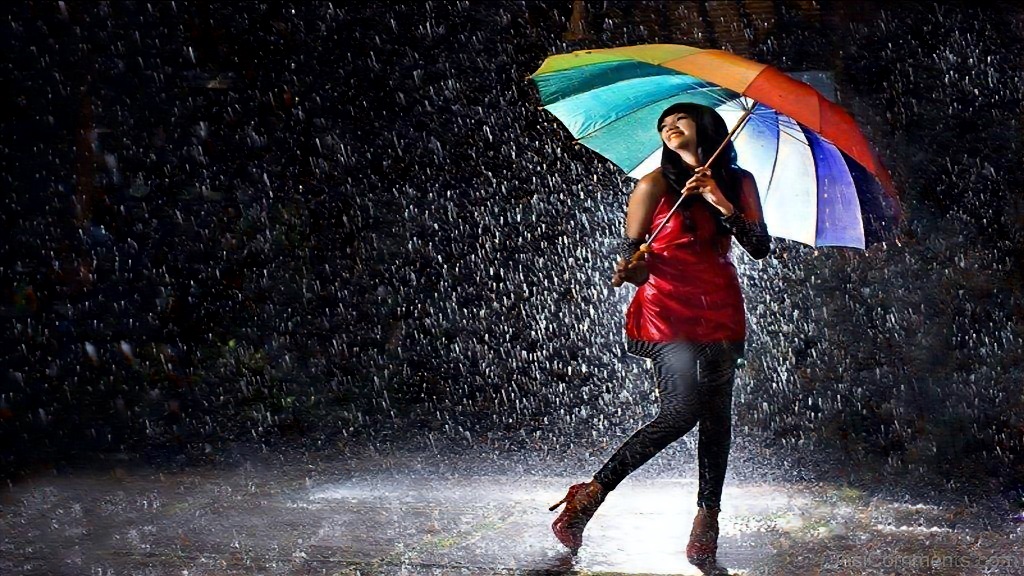}  
		\vspace*{-0.5cm}
		\subcaption*{SEMI \cite{semi} \\ 0.1733}
	\end{subfigure}
	\hspace{-0.12cm}
	\begin{subfigure}[t]{0.2\linewidth}
		\includegraphics[width=1.\linewidth]{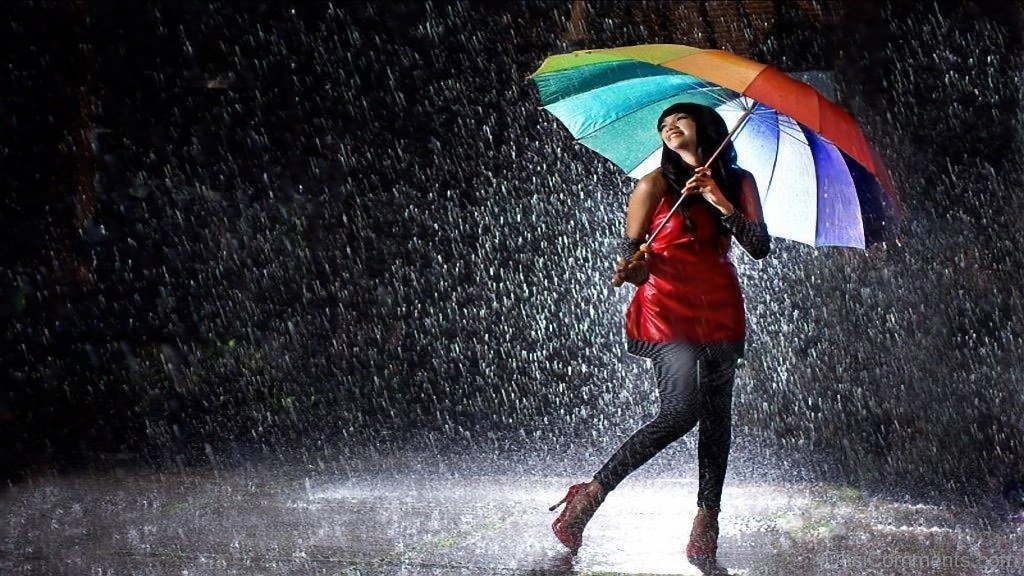}  
		\vspace*{-0.5cm}
		\subcaption*{SPANet \cite{spanet} \\ 0.3933}
	\end{subfigure}
	\hspace{-0.12cm}
	\begin{subfigure}[t]{0.2\linewidth}
		\includegraphics[width=1.\linewidth]{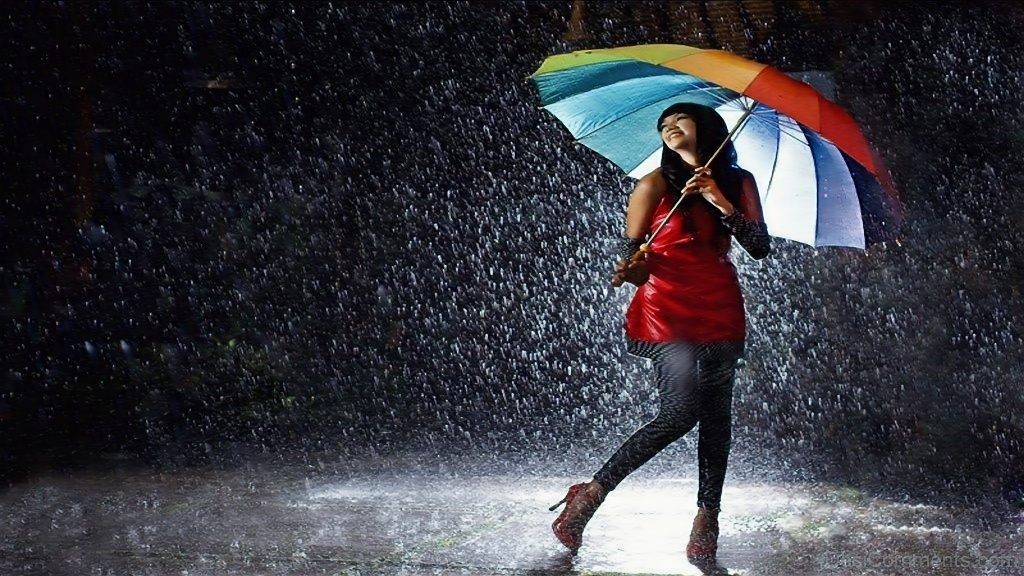}  
		\vspace*{-0.5cm}
		\subcaption*{PreNet \cite{prenet} \\ 0.1883}
	\end{subfigure}
	\hspace{-0.12cm}
	\begin{subfigure}[t]{0.2\linewidth}
		\includegraphics[width=1.\linewidth]{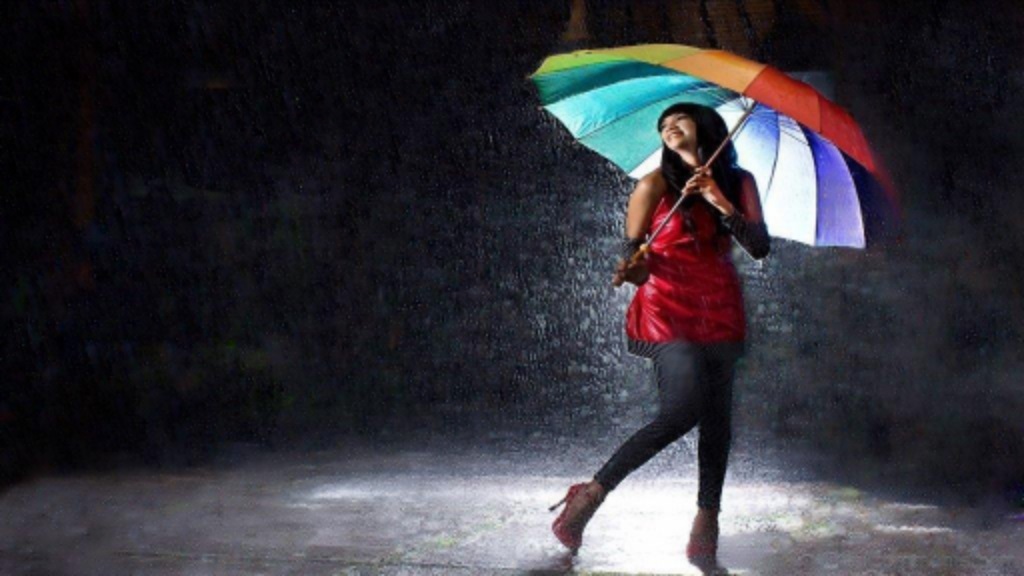}  
		\vspace*{-0.4cm}
		\subcaption*{CODE-Net (ours) \\0.1050}
	\end{subfigure}
	\hspace{-0.12cm}
	\begin{subfigure}[t]{0.2\linewidth}
		\includegraphics[width=1.\linewidth]{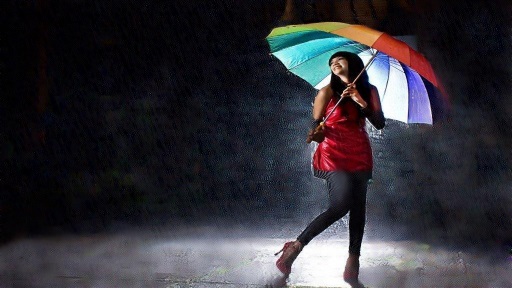}  
		\vspace*{-0.4cm}
		\subcaption*{mCODE-Net (ours) \\0.0900}
	\end{subfigure}
	\caption{Qualitative comparisons of two real images with heavy rain and the corresponding RDEs are calculated by \myref{rain-density-estimation}.}
	\label{Qualitative_comparison_on_real}
\end{figure*}

\begin{figure*}[!t]
	\centering
	\begin{subfigure}[t]{0.2\linewidth}
		\includegraphics[width=1.\linewidth]{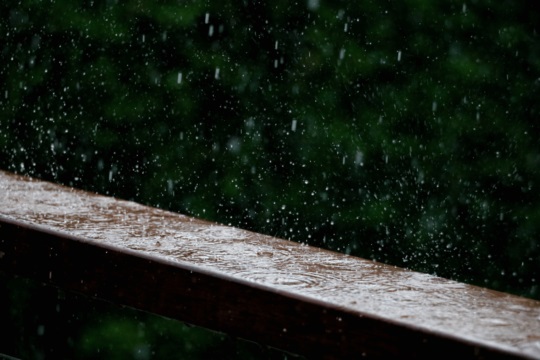} 
		\vspace*{-0.5cm}
		\subcaption*{Input \\ 0.0183}
	\end{subfigure}
	\hspace{-0.12cm}
	\begin{subfigure}[t]{0.2\linewidth}
		\includegraphics[width=1.\linewidth]{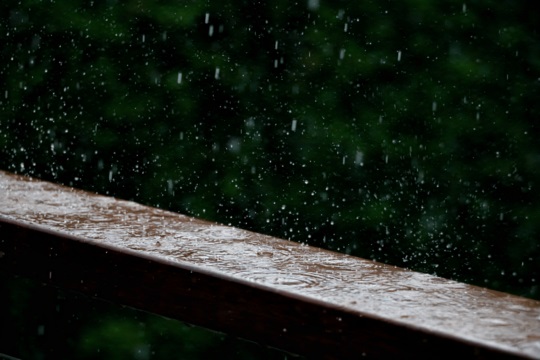} 
		\vspace*{-0.5cm}
		\subcaption*{DDN \cite{ddn} \\ 0.0067}
	\end{subfigure}
	\hspace{-0.12cm}
	\begin{subfigure}[t]{0.2\linewidth} 
		\includegraphics[width=1.\linewidth]{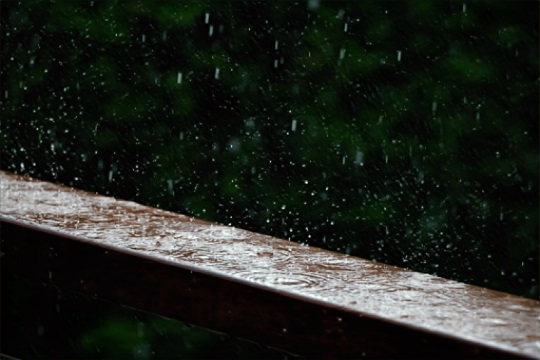}  
		\vspace*{-0.5cm}
		\subcaption*{DID-MDN \cite{did-mdn} \\0.0250}
	\end{subfigure} 
	\hspace{-0.12cm}
	\begin{subfigure}[t]{0.2\linewidth}
		\includegraphics[width=1.\linewidth]{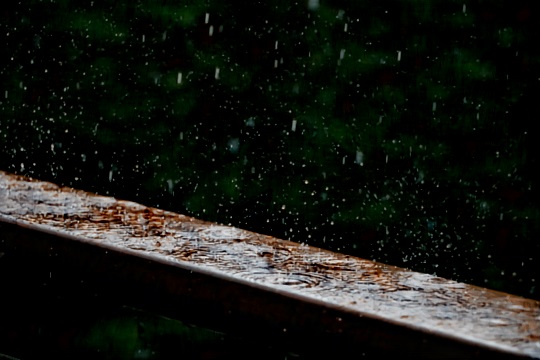}  
		\vspace*{-0.5cm}
		\subcaption*{SEMI \cite{semi} \\ 0.0083}
	\end{subfigure}
	\hspace{-0.12cm}
	\begin{subfigure}[t]{0.2\linewidth}
		\includegraphics[width=1.\linewidth]{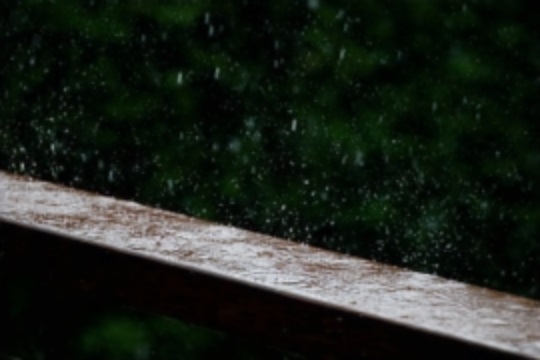}  
		\vspace*{-0.5cm}
		\subcaption*{SPANet \cite{spanet} \\0.0066}
	\end{subfigure}
	\hspace{-0.12cm}
	\begin{subfigure}[t]{0.2\linewidth}
		\includegraphics[width=1.\linewidth]{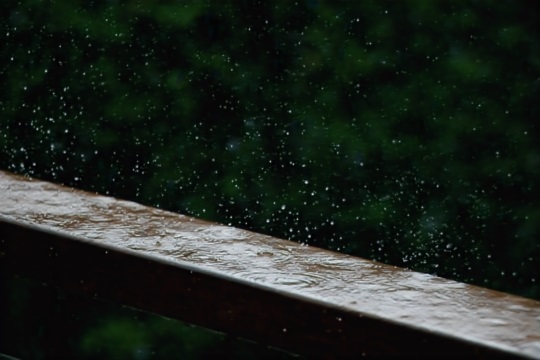}  
		\vspace*{-0.5cm}
		\subcaption*{PreNet \cite{prenet} \\0.0133}
	\end{subfigure}
	\hspace{-0.12cm}
	\begin{subfigure}[t]{0.2\linewidth}
		\includegraphics[width=1.\linewidth]{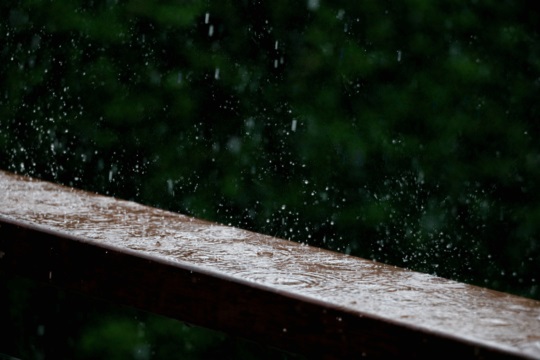}  
		\vspace*{-0.4cm}
		\subcaption*{CODE-Net (ours) \\0.0150}
	\end{subfigure}
	\hspace{-0.12cm}
	\begin{subfigure}[t]{0.2\linewidth}
		\includegraphics[width=1.\linewidth]{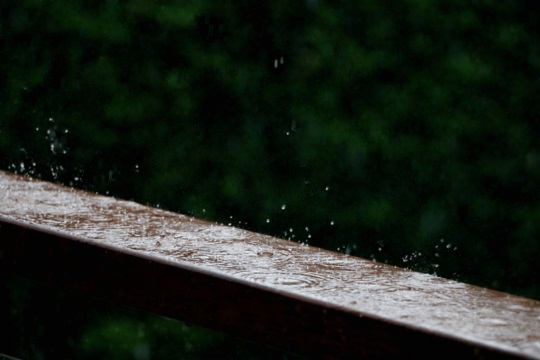}  
		\vspace*{-0.4cm}
		\subcaption*{mCODE-Net (ours) \\0.0100}
	\end{subfigure}\\	
	\centering
	\begin{subfigure}[t]{0.2\linewidth}
		\includegraphics[width=1.\linewidth]{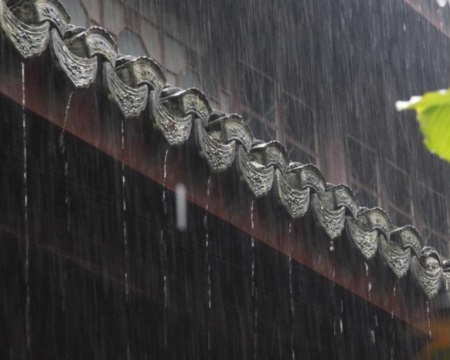} 
		\vspace*{-0.5cm}
		\subcaption*{Input \\0.3216}
	\end{subfigure}
	\hspace{-0.12cm}
	\begin{subfigure}[t]{0.2\linewidth}
		\includegraphics[width=1.\linewidth]{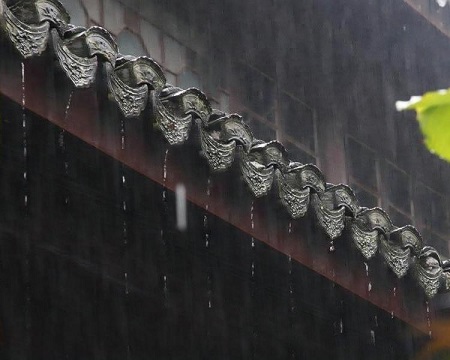} 
		\vspace*{-0.5cm}
		\subcaption*{DDN \cite{ddn} \\0.1350}
	\end{subfigure}
	\hspace{-0.12cm}
	\begin{subfigure}[t]{0.2\linewidth} 
		\includegraphics[width=1.\linewidth]{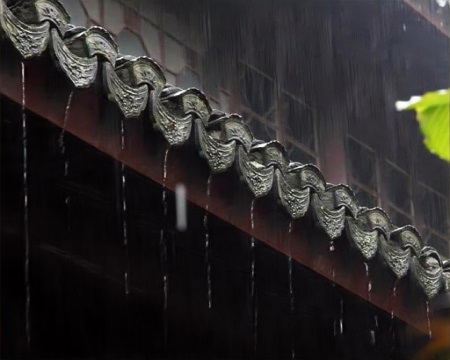}  
		\vspace*{-0.5cm}
		\subcaption*{DID-MDN \cite{did-mdn} \\0.1600}
	\end{subfigure} 
	\hspace{-0.12cm}
	\begin{subfigure}[t]{0.2\linewidth}
		\includegraphics[width=1.\linewidth]{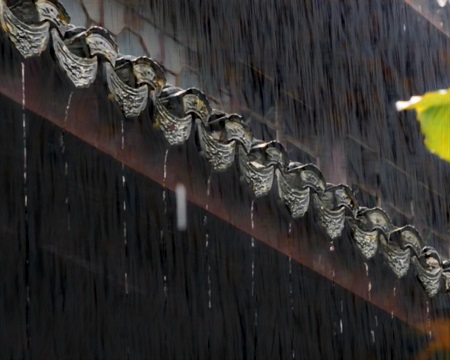}  
		\vspace*{-0.5cm}
		\subcaption*{SEMI \cite{semi} \\0.3017}
	\end{subfigure}
	\hspace{-0.12cm}
	\begin{subfigure}[t]{0.2\linewidth}
		\includegraphics[width=1.\linewidth]{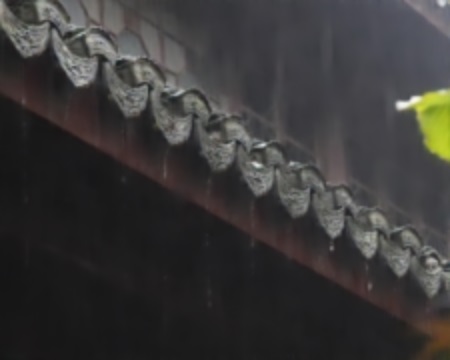}  
		\vspace*{-0.5cm}
		\subcaption*{SPANet \cite{spanet} \\0.0800}
	\end{subfigure}
	\hspace{-0.12cm}
	\begin{subfigure}[t]{0.2\linewidth}
		\includegraphics[width=1.\linewidth]{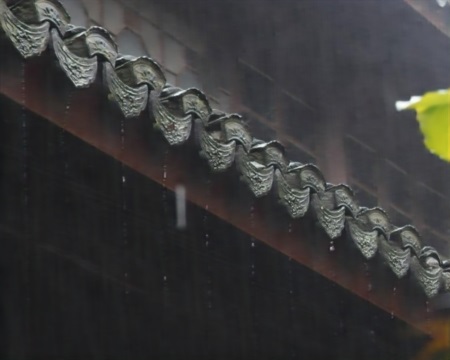}  
		\vspace*{-0.5cm}
		\subcaption*{PreNet \cite{prenet} \\ 0.1017}
	\end{subfigure}
	\hspace{-0.12cm}
	\begin{subfigure}[t]{0.2\linewidth}
		\includegraphics[width=1.\linewidth]{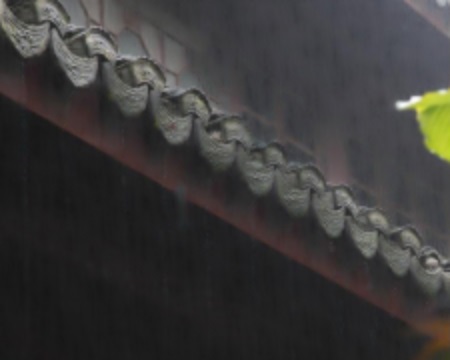}  
		\vspace*{-0.4cm}
		\subcaption*{CODE-Net (ours) \\ 0.0717}
	\end{subfigure}
	\hspace{-0.12cm}
	\begin{subfigure}[t]{0.2\linewidth}
		\includegraphics[width=1.\linewidth]{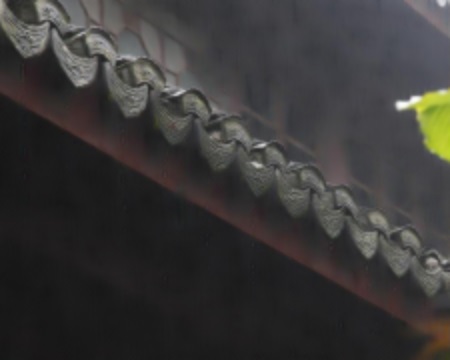}  
		\vspace*{-0.4cm}
		\subcaption*{mCODE-Net (ours) \\0.0683}
	\end{subfigure}
	\caption{Qualitative comparisons of two real images with light rain and the corresponding RDEs are calculated by \myref{rain-density-estimation}.}
	\label{Qualitative_comparison_on_real1}
\end{figure*}
	\subsection{Network Architecture}
	In this section, we will introduce the network structure of CODE-Net, mainly consisting of rain streaks extractor and rain streaks denoiser.
	
	\textbf{{Rain Streaks Extractor.}} We use two convolutional layers $\boldsymbol{E}_{1} $, $ \boldsymbol{E}_{2}$  to extract noisy rain streaks $ \boldsymbol{r}_{\epsilon} $ from the rainy input $ \boldsymbol{y} $:
	\begin{equation}
	\boldsymbol{r}_{\epsilon}  =\text{ReLU} \left( \boldsymbol{E}_{2} \otimes \text{ReLU} \left( \boldsymbol{E}_{1}\otimes\boldsymbol{y}\right)\right),
	\end{equation}
	as shown in Fig.~\ref{fig:model}, $ \boldsymbol{r}_{\epsilon} $ can be seen as consisting of rain streaks and noise.
	
	\textbf{{Rain Streaks Denoiser.}} $ \boldsymbol{r}_{\epsilon} $ is then fed into  \myref{rewrite_final_solution_to_reweight_csc} to seek for the convolutional sparse code $ \boldsymbol{z}$. To avoid long inference time, we build \myref{rewrite_final_solution_to_reweight_csc} by CNN techniques.
	
	Specifically, for $ \Gamma_{\theta}(\cdot) $, under the assumption of nonnegative sparse coding, the soft nonnegative thresholding operator is equivalent to ReLU \cite{papyan2017convolutional}. Thus, in \myref{rewrite_final_solution_to_reweight_csc}, we replace $ \Gamma_{\theta}\left( \alpha \right)  $ with $ \text{ReLU}\left( \alpha-\theta \right)  $, where the threshold $\theta$ is learnable as well. Even though \myref{equ:final-lwista} is for one channel, the extension to multichannel
	case is mathematically straightforward. That is, for an input with $p$ channels, \myref{equ:final-lwista} still holds true with $\boldsymbol{G} \in \mathbb{R}^{c\times p\times s\times s}$, $\boldsymbol{S} \in \mathbb{R}^{c\times c\times s\times s}$, which means $\boldsymbol{S}$ and $ \boldsymbol{G} $ correspond to two convolution layers. Above all, the resulting learning weighted ISTA (LwISTA) can be expressed as:
	\begin{equation} \label{equ:final-lwista}
	\begin{split}
	\boldsymbol{v}^{(t)} &=\boldsymbol{S} \otimes \boldsymbol{z}^{(t)} + \boldsymbol{G} \otimes \boldsymbol{r}_\epsilon  \\
	\boldsymbol{z}^{(t+1)}&=\text{LWB}_{\boldsymbol{w}} \left(  \text{ReLU}\left(\text{LWB}_{\tilde{\boldsymbol{w}}}  \left( \boldsymbol{v}^{(t)}\right)  - \theta\right)\right).
	\end{split}
	\end{equation}
	
	Based on \myref{equ:final-lwista}, initializing $ \boldsymbol{z}^{(0)} $ to $ \boldsymbol{0} $, the optimal convolutional sparse code  could be produced efficiently after $ T $ iterations. Note that $ \boldsymbol{S} $ shares the parameters over layers, as well as $ \text{LWB}_{ \boldsymbol{w} /\tilde{\boldsymbol{w}}} $.
	
	After obtaining the convolutional sparse code $ \boldsymbol{z} $, to recover the noiseless rain streak $\boldsymbol{r} $, another dictionary $\boldsymbol{E}_{3} $ corresponding to $ \boldsymbol{G} $ and a convolutional layer $ \boldsymbol{E}_{4}$ are needed:
	\begin{equation}
	\boldsymbol{r}  = \boldsymbol{E}_{4} \otimes \text{ReLU} \left( \boldsymbol{E}_{3}\otimes\boldsymbol{z}\right).
	\end{equation}
	
	Finally, the clean background image $ \boldsymbol{x} $ would be produced by subtracting rain streaks $ \boldsymbol{r} $ from the rainy image $ \boldsymbol{y} $:
	\begin{equation}
	\boldsymbol{x}  = \boldsymbol{y} -  \boldsymbol{r}.
	\end{equation}
	The whole architecture is illustrated in Fig. \ref{fig:model}.

	\vspace{-0.1cm}
	\section{Multiscale CODE-Net: mCODE-Net}
	In deraining task, it is well studied in \cite{mscsc} that multiscale CSC performs better than traditional CSC  since the rain streaks, captured from different distances, always appear to be multiscale. Inspired by this, we further extend CODE-Net to its multiscale version (mCODE-Net) to  mine the potential,  by expanding \myref{eq:2} to the following (three scales):
	\begin{equation}\label{eq:mCODE-Net-rain}
	\boldsymbol{r}_\epsilon = \sum_{j=1}^{3}\sum_{i=1}^{c} \boldsymbol{f}_{j,i} \otimes \boldsymbol{z}_{j,i} + \boldsymbol{\epsilon},
	\end{equation}
	where $\left\{\boldsymbol{f}_{j,i}\right\}_{i=1}^{c} \in \mathbb{R}^{c \times s_j \times s_j} $ and $\left\{\boldsymbol{z}_{j,i}\right\}_{i=1}^{c} \in \mathbb{R}^{c \times l_h \times l_w} $ are rain filters of $s_j \times s_j$ and representations corresponding to the $j$-th dictionary.
	
	Analogy to the solution procedure in Section \ref{denoising_by_lwista},  one can obtain the convolutional sparse  codes for the $j$-th dictionary by:
	\begin{equation} \label{equ:final-lwista-ms}
	\begin{split}
	\boldsymbol{v}^{(t)}_{j} &= \boldsymbol{z}^{(t)}_{j} + \boldsymbol{G}_j \otimes  \left( \boldsymbol{r}_\epsilon -  \sum_{m=1}^{3} \boldsymbol{S}_m  \otimes \boldsymbol{z}_m^{(t)}  \right)  \\
	\boldsymbol{z}^{(t+1)}_{j} &=\text{LWB}_{{w},j} \left(  \text{ReLU}\left(\text{LWB}_{\tilde{w},j}  \left( \boldsymbol{v}^{(t)}_{j}\right)  - \theta_{j}^{(t)}\right)\right),
	\end{split}
	\end{equation}
	and recover the noiseless rain streak $\boldsymbol{r} $ by:
	\begin{equation}
	\boldsymbol{r}  = \boldsymbol{E}_{4} \otimes \text{ReLU} \left( \boldsymbol{S}_1  \otimes \boldsymbol{z}_1 +\boldsymbol{S}_2  \otimes \boldsymbol{z}_2 + \boldsymbol{S}_3  \otimes \boldsymbol{z}_3 \right),
	\end{equation}
	where $\left\{ \boldsymbol{G}_j \in \mathbb{R}^{c \times p \times s_j \times s_j}, \boldsymbol{S}_j \in \mathbb{R}^{p \times c \times s_j \times s_j} \right\}_{j=1}^{3} $ can be implemented by several  convolutional layers of $s_j \times s_j$. $ \text{LWB}_{{w},j}  $/$\text{LWB}_{\tilde{w},j}$ and $ \theta_{j} $ are learning weight blocks and threshold \textit{w.r.t.} the $j$-th dictionary.

	\section{Experimental Results}
	\label{sec:exper}
	\subsection{Settings}
	\textbf{Datasets and Metrics.} {\color{black}We use 12000 and 1800  pairs of images from \cite{did-mdn,jorder} and a high quality real rain dataset~\cite{spanet} as { the} training set. For testing, three commonly synthetic datasets, Rain1200 \cite{did-mdn}, Rain12 \cite{gmm} and Test1000~\cite{spanet}, and some real-world images are utilized.} The Rain1200 testing set, consisting of heavy, medium and light rain (Fig.~\ref{fig:example-rain1200}), is more challenging. Deraining results on synthetic datasets are evaluated with Peak Signal-to-Noise Ratio (PSNR) and Structural Similarity (SSIM) \cite{ssim} on the Y channel of transformed YCbCr space, while the performance of real-world images is evaluated visually since the ground truth images are not available.
	
	\textbf{Model.} In CODE-Net, each convolutional layer has $ 128 $ filters ($p=128$) of { size} $3 \times 3$ ($s=3$) except $ \boldsymbol{G} $, $ \boldsymbol{S} $ and $\boldsymbol{E}_4$ { respectively} having $ 256 $, $256$ ($c=256$) and $3$ filters. By contrast, we set $p=48$, $ c=96$, and $s_1=3$, $ s_2=5$, $ s_3=7$ (namely, the rain filters are of size $3\times 3$, $5\times 5$ and $7\times 7$)  in mCODE-Net\footnote{{Note that, even though mCODE-Net is multiscale, it has fewer parameters ($802k$ \textit{vs.} $1350k$) than CODE-Net since fewer filters { are exploited}.}}. We desert the biases and use zero-padding to keep  the size of feature maps fixed. {\color{black} Another important point is, differ from \cite{ddn,did-mdn}, there is no BatchNormalization \cite{bnlayers} in our models according to \myref{equ:final-lwista} and \myref{equ:final-lwista-ms}. 
	
	\textbf{Training.} To make full use of the limited images, we augment the training set with random horizontal flips and $ 90 $ degree rotations. The training process is divided into two stages: (1)  we train two plain models (CODE-Net/mCODE-Net without weighting), initialized by the method of He \textit{et al.}~\cite{He2015DelvingDI}, using $\ell_{1}$ loss in RGB channels. For optimization, we use Adam \cite{adam} with $\beta_1=0.9$, $\beta_2=0.999$. In each training batch, 8 patches of size $128 \times 128$ are extracted as inputs. The learning rate is set to $8 \times 10^{-4}$ and halved at $[50k$, $100k$, $150k$, $200k]$ iterations; (2) we then initialize CODE-Net and mCODE-Net using the plain models and fine-tune them with the learning rate set to $8 \times 10^{-5}$ and halved at $[50k$, $100k$, $150k$, $200k]$ iterations. We use PyTorch  to implement our models with an NVIDIA RTX 2080 GPU\footnote{{Codes and more results are available at \href{https://github.com/Achhhe/CODE-Net}{https://github.com/Achhhe/CODE-Net}}}.

	\subsection{Density Estimation of Synthetic and Real Rain}
	\label{subsec:rain-density-estimation}
	{\color{black}
		In this subsection, we focus on the rain density estimation of rainy images with synthetic and real rain. For synthetic rainy images, we show a sample of different rain levels in Fig.~\ref{fig:example-rain1200}. And in Fig.~\ref{fig:analysis-weight}, RDEs of several samples are depicted.  One can see that, the RDEs mainly range from $0.1$ to $0.95$ and become larger as the {\color{black}rain density} increases, consistent with the analysis in Section \ref{rde-with-weights}. 
		
		Besides, a clear image and some real-world rainy images with continuous rainy states and their RDEs are shown in Fig.~\ref{fig:real-rie-all}. The clear image has the smallest RDE $ 0.1617 $ while the rainy images have  gradual increased RDEs from $0.2550$ to $0.9850$, which verifies the validity of continuous rain density estimation of our model.	 
	}
	
	\begin{table}[!t]  
		\centering  
		\fontsize{7.5}{9}\selectfont  
		\begin{threeparttable}  
			\caption{Average PSNR/SSIMs on Rain12, Rain1200 and Test1000.  \textbf{Bold}: the best; \underline{underline}: the second best.} 
			
			\label{tab:performance_comparison}
			\begin{tabular}{cccc}
				\toprule  
				\multirow{2}{*}{\makecell{Methods\\[-.15cm]}}& 
				\multicolumn{1}{c}{Rain12}&\multicolumn{1}{c}{Rain1200}&\multicolumn{1}{c}{Test1000}\cr  
				\cmidrule(lr){2-2} \cmidrule(lr){3-3}\cmidrule(lr){4-4}   
				&PSNR/SSIM&PSNR/SSIM&PSNR/SSIM\cr  
				\midrule 
				
				DSC*\cite{dsc}  &29.98/0.8654  &21.44/0.7896&32.33/0.9305\cr  
				GMM*\cite{gmm} &32.15/0.9145&22.75/0.8352&32.99/0.9475\cr 
				CNN~\cite{cnn}   &33.33/0.9199&23.55/0.8352&31.31/0.9304\cr
				JORDER~\cite{jorder}   & {36.15}/0.9548&25.71/0.8074&35.72/{0.9776}\cr
				DDN~\cite{ddn}   & 29.84/0.9049& 30.08/0.8791&34.88/0.9727\cr 
				DID-MDN~\cite{did-mdn} &29.49 /0.9031&29.65/0.9016&28.96/0.9457\cr
				PreNet~\cite{prenet} & \underline{36.66} /0.9610&30.56/0.8750&30.31/0.9538\cr
				SPANet~\cite{spanet} &32.71 /0.9285&30.05/\bf 0.9342&38.53/ \underline{0.9875}\cr
				\midrule
				CODE-Net (ours) &{ 36.21}/\underline{ 0.9618}&{ \underline {33.31} }/0.9174&\underline{ 38.88}/0.9867\cr
				mCODE-Net (ours) &\bf { 36.79}/{ \bf {0.9639}}&{\bf {34.03}}/\underline{ 0.9281}&{\bf 39.85}/\bf 0.9879\cr  				
				\bottomrule  
			\end{tabular} 
			
		\end{threeparttable}  
	\end{table}  
	
	\subsection{Deraining Result Comparison}
	\label{derain-comparison}
	In this subsection, we compare our proposed CODE-Net and mCODE-Net to 9 state-of-the-art deraining methods, including DSC \cite{dsc}, GMM~\cite{gmm}, CNN~\cite{cnn}, DDN~\cite{ddn}, JORDER~\cite{jorder}, DID-MDN~\cite{did-mdn}, SEMI~\cite{semi}, SPANet~\cite{spanet} and PreNet~\cite{prenet}. All results are reproduced by the authors' codes except the quantitative ones of DSC and GMM on Rain1200 are cited from \cite{did-mdn} directly since their codes run slowly.

	\textbf{Synthetic Images.} 
	Quantitative results are tabulated in Tab.~\ref{tab:performance_comparison}. Clearly, both CODE-Net and mCODE-Net achieve superior performance. {Especially on Rain1200, the more complex dataset consisting of light, medium and heavy levels, our models outperform others by large margins, which indicates our models are more robust for different densities of rain. One important reason is that the rain density is implicitly considered in \myref{rewrite_final_solution_to_reweight_csc} and thus leads to better rain removal results.} Fig.~\ref{Qualitative_comparison_on_synthetic_rain12} and Fig.~\ref{Qualitative_comparison_on_synthetic_rain1200} show the visual comparisons. Our models consistently achieve the best visual performance in terms of effectively removing rain streaks
	while preserving background details. Besides, thanks to the multiscale prior, mCODE-Net could achieve comparable or even better results than CODE-Net with fewer parameters. 
	
	Note that, JORDER and DID-MDN use additional rain mask or density information in training stage. Nonetheless, our models still greatly outperform their results without any additional data.

	\textbf{Real Images.} To demonstrate the generalization of the proposed architectures, we evaluate our models and other methods on four real-world images, shown in  Fig.~\ref{Qualitative_comparison_on_real} and Fig.~\ref{Qualitative_comparison_on_real1}. Once again, our models show the best visual performance. In contrast, previous methods especially DID-MDN tend to produce under/over deraining results. Our models get rid of these faults and produce more pleasant results, which verifies the effectiveness of continuous rain density estimation over classifying it into three levels. 
	
	{ Comparing CODE-Net and mCODE-Net, raindrops with large or small size can be efficiently removed by mCODE-Net with further consideration of the multi-scale property, while CODE-Net gives a result by leveraging different raindrop scales, as shown in Fig.~\ref{Qualitative_comparison_on_real}.}
	
	{ We further apply the CODE-Net on the deraining results obtained by each method and use RDE \myref{rain-density-estimation} to estimate their rain densities. As shown in Fig.~\ref{Qualitative_comparison_on_real} and Fig.~\ref{Qualitative_comparison_on_real1}, the RDEs of the deraining images decrease comparing to the original rainy inputs, which reflects the deraining effects of each method. One can find that the proposed CODE-Net and mCODE-Net can achieve much lower RDEs than other methods for rainy images with heavy rains shown in Fig.~\ref{Qualitative_comparison_on_real}. It  coincides with the results on the synthetic rainy images as shown in Tab.~\ref{tab:performance_comparison}, where CODE-Net and mCODE-Net can outperform the state-of-the-arts by large margins. 
	
	{On the other hand}, the RDEs obtained from CODE-Net have potentials to quantitatively evaluate the deraining performance for rainy images from real scenes.}

	\begin{figure}[!t]
	\centering
	\includegraphics[width=0.9\linewidth]{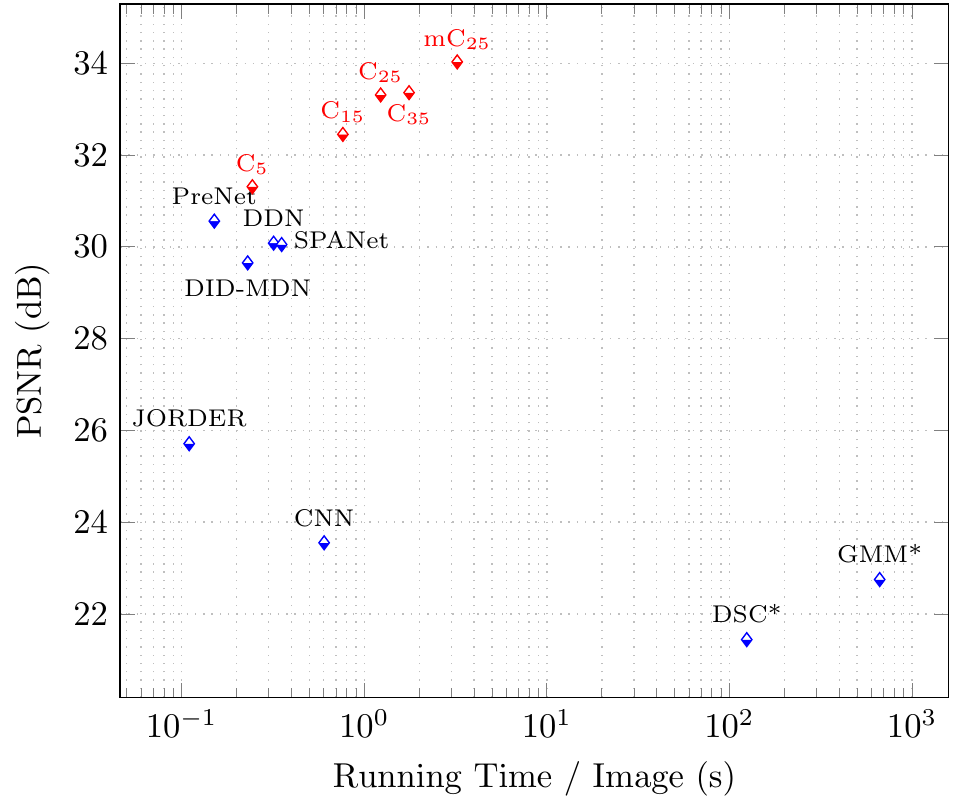}
	\caption{Performance of state-of-the-art methods versus the running time on Rain1200. `*' indicates that the method is tested on CPU. C$ _T $/mC$ _T $ denote CODE-Net/mCODE-Net iterating $ T $ times. The results show that our models could balance performance and running time by changing $ T $. }
	\label{fig:runtime}
\end{figure}
	
	\subsection{Running Time Comparison}
	{ We visualize the running time per image ($ 512 \times 512 $) versus performance (PSNR) on Rain1200 dataset in Fig.~\ref{fig:runtime}. According to the provided source code, all methods are tested on GPU except DSC and GMM on CPU. One key factor affecting running time of our method is the number of iterations $ T $ when solving convolutional sparse codes in \myref{equ:final-lwista}. For simplicity, we denote CODE-Net/mCODE-Net iterating $ T $ times as C$ _T $/mC$ _T $. It is observed that C$ _5 $ achieves comparable performance and computational complexity as other {\color{black}state-of-the-art methods}. However, too few iterations are not capable to generate {\color{black}good} convolutional sparse codes. Hence, we additionally test three models with $ T=15, 25, 35 $ and unsurprisingly the results of them  outperform those of other methods largely.
	
	 As shown in Fig.~\ref{fig:runtime}, with the increase of the number of iterations, the performance of our method can be {\color{black}further} improved and {\color{black}obtain the optimal result} when $ T=25 $. In this paper, we focus on the best rain removal performance and thus choose $ T=25 $ in our final models (C$_{25}$/mC$_{25}$). For practical applications, we could adjust the number of iterations $ T $ to balance performance and running time. }

	\subsection{{Ablation Study}}
	{Our network (Fig.~\ref{fig:model}) is derived from traditional CSC problem, but by coincidence, it includes two common strategies in deep learning, \textit{i.e.}, local residual learning (\textbf{LRL})~\cite{drrn} and pre-activation (\textbf{PA})~\cite{he2016identity}. LRL is actually the local skip connection and PA means activation layer (LWB-ReLU-LWB) comes before weight layer ($\boldsymbol{S}$) in unfolded LwISTA. Besides, to improve the deraining performance, we introduce {reweighting} (\textbf{RW}) into CSC optimization. Last but not the least, since their specific characteristics, we focus on extracting rain streaks and  recover the deraining image by  subtracting them from the input. Interestingly, this  corresponds to the global residual learning (\textbf{GRL}). }
	
	\begin{figure}[!t] 
	\raggedleft
	\begin{subfigure}[t]{0.48\linewidth}
		\includegraphics[width=0.95\linewidth]{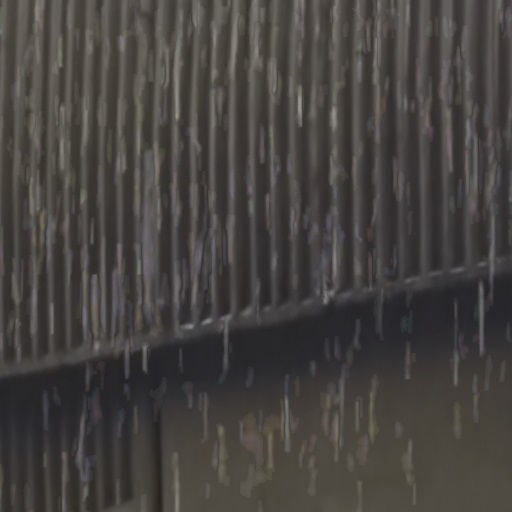}
		\subcaption{RDE: 0.6550}
		\label{fig:failure3}
	\end{subfigure}
	\raggedright
	\begin{subfigure}[t]{0.48\linewidth}
		\includegraphics[width=0.95\linewidth]{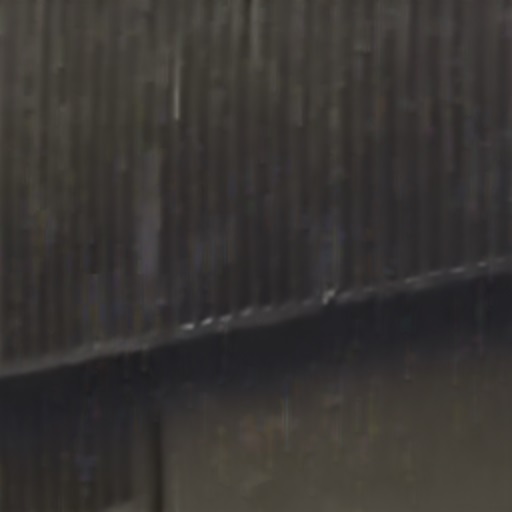}
		\subcaption{RDE: 0.1023 }
		\label{fig:failure4}
	\end{subfigure} \\
	\raggedleft
	\begin{subfigure}[t]{0.48\linewidth}
		\includegraphics[width=0.95\linewidth]{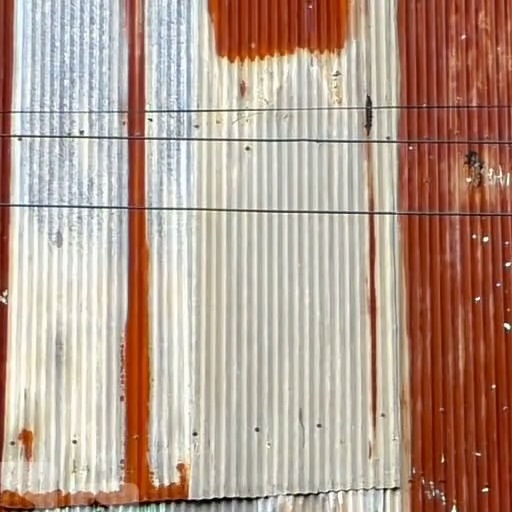}
		\subcaption{RDE: 0.8923}
		\label{fig:failure1}
	\end{subfigure}
	\raggedright
	\begin{subfigure}[t]{0.48\linewidth}
		\includegraphics[width=0.95\linewidth]{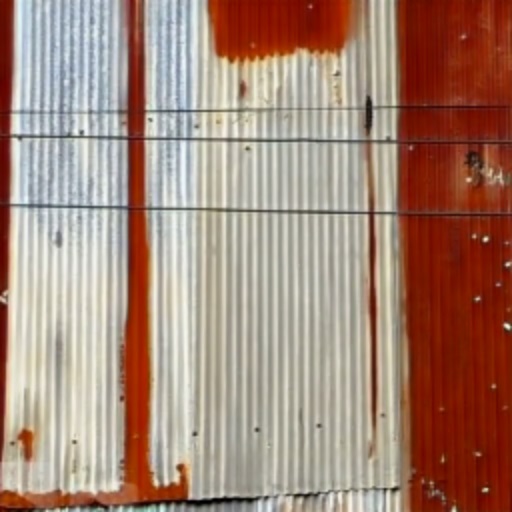}
		\subcaption{ RDE: 0.7416}
		\label{fig:failure2}
	\end{subfigure} 
	\caption{ Failure cases of the proposed method.  Rainy inputs (a)\&(c) and deraining results (b)\&(d), with the corresponding RDEs. The proposed method may be confused by images with rain-like structures and produce overestimations of RDEs and over deraining results.}
	\label{fig:failure}
\end{figure}
	\begin{figure*}[!t] 
	\begin{subfigure}[t]{0.99\linewidth}
		\begin{minipage}[b]{1\linewidth}
			\begin{subfigure}[t]{0.24\linewidth}
				\includegraphics[width=1\linewidth]{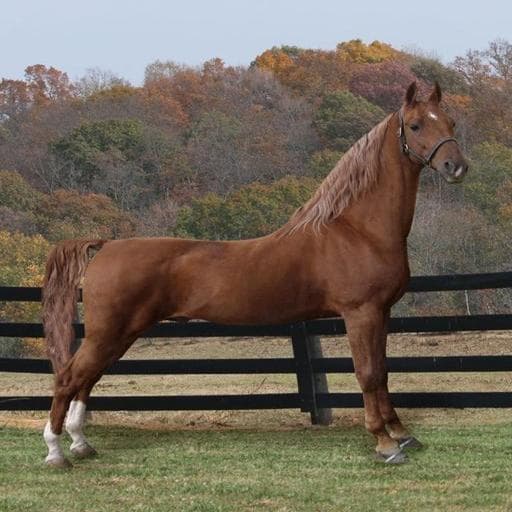}
				\vspace*{-0.5cm}
				\subcaption*{GT \\ (PSNR, SSIM) }
				\label{fig:ext-esrgan-gt}
			\end{subfigure}
			\hspace{-0.112cm}
			\begin{subfigure}[t]{0.24\linewidth}
				\includegraphics[width=1\linewidth]{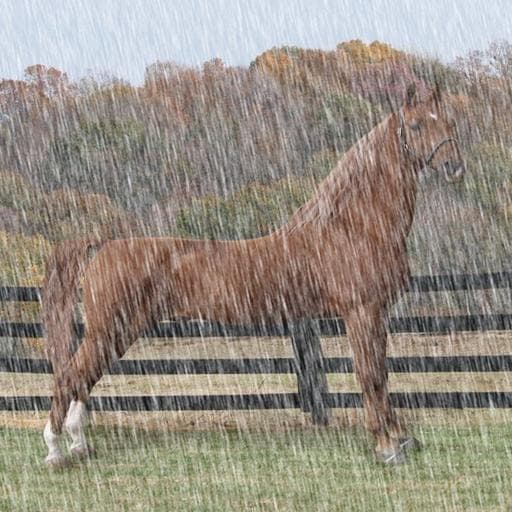}
				\vspace*{-0.5cm}
				\subcaption*{Rainy input \\ (16.67, 0.5138)}
				\label{fig:ext-esrgan-inp}
			\end{subfigure}
			\hspace{-0.112cm}
			\begin{subfigure}[t]{0.24\linewidth}
				\includegraphics[width=1\linewidth]{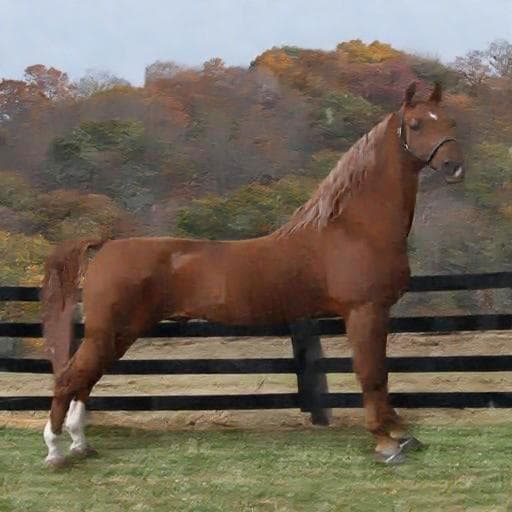}
				\vspace*{-0.5cm}
				\subcaption*{CODE-Net \\ (29.85, 0.8112)}
				\label{fig:ext-esrgan-psnr}
			\end{subfigure}
			\hspace{-0.112cm}
			\begin{subfigure}[t]{0.24\linewidth}
				\includegraphics[width=1\linewidth]{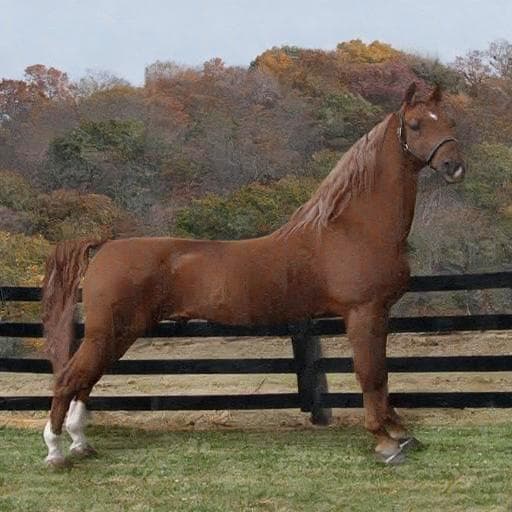}
				\vspace*{-0.5cm}
				\subcaption*{GaCODE-Net \\ (29.10, 0.8083)}
				\label{fig:ext-esrgan-gan}
			\end{subfigure}
		\end{minipage}
		\subcaption{{Visual pursuit. By using perceptual loss, GaCODE-Net could recover more realistic textures and high-frequency details.}}
		\label{fig:extesions-visual-pursuit}
	\end{subfigure}
	\begin{subfigure}[t]{0.99\linewidth}
	\vspace{0.2cm}
	\begin{minipage}[b]{1\linewidth}
			\begin{subfigure}[t]{0.49\linewidth}
			\includegraphics[width=.5\linewidth]{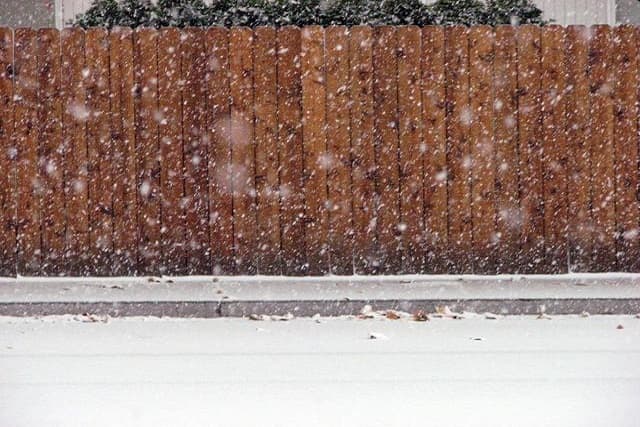}\includegraphics[width=.5\linewidth]{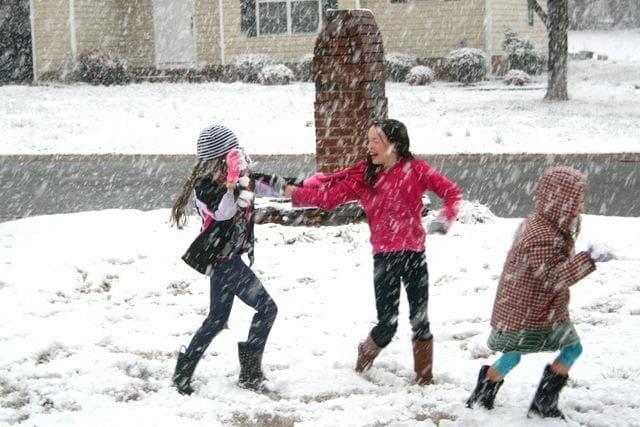}
			\vspace*{-0.5cm}
			\subcaption*{Snowy inputs}
		\end{subfigure}
		\hspace{-0.112cm}
		\begin{subfigure}[t]{0.49\linewidth}
			\includegraphics[width=.5\linewidth]{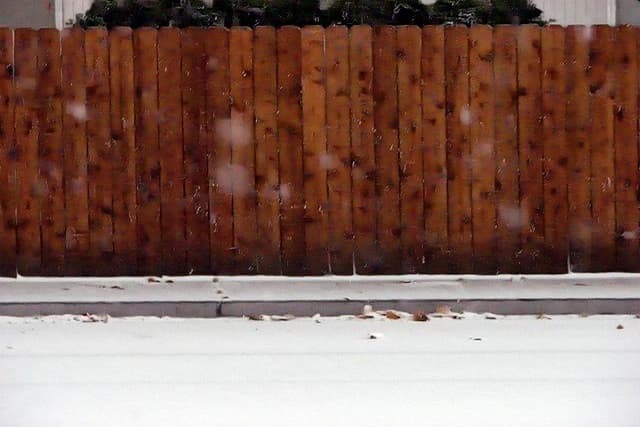}\includegraphics[width=.5\linewidth]{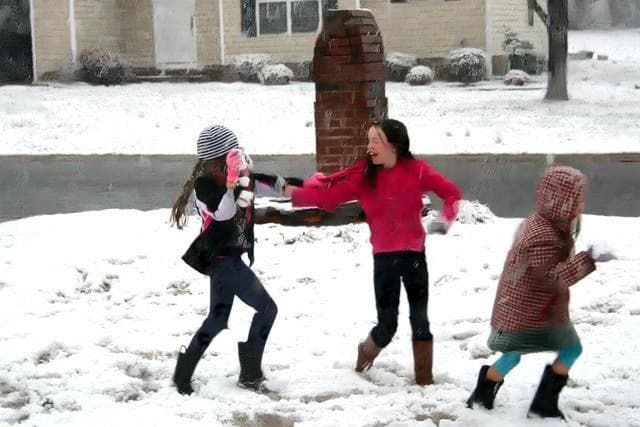}
			\vspace*{-0.5cm}
			\subcaption*{Desnowing results}
			\label{fig:snow-results}
		\end{subfigure}
	\end{minipage}
	\subcaption{{Single image  desnowing. The results show that our CODE-Net has excellent performance in the snow removal task.}}
\label{fig:extesions-desnowing}
\end{subfigure}
	\caption{{The generalization of our CODE-Net to other tasks, including visual pursuit via perceptual loss and image desnowing.} }
	\label{fig:extesions}
\end{figure*}

	To demonstrate the effectiveness of aforementioned components, we conduct an ablation study of GRL, LRL, {PA} and RW on CODE-Net.  $ 96 $ images are selected from Rain1200 for testing, with each level $ 32 $ images. Tab.~\ref{tab:ablation_study} shows the average PSNR. When both LRL and PA are used, the PSNR is relatively high, no matter RW is used or not (\nth{2}/\nth{3} \textit{vs.} \nth{8}/\nth{9}). Comparing \nth{2}/\nth{4}/\nth{6}/\nth{8} with \nth{3}/\nth{5}/\nth{7}/\nth{9}, one can see that the performance could be improved by adding RW. However, it is interesting that, using PA only would degrade the performance (\nth{2}/\nth{3} \textit{vs.} \nth{4}/\nth{5}). Not surprisingly, CODE-Net using four components simultaneously performs the best. These
	comparisons indicate the superiority of reweighted CSC inspired network combining GRL, LRL, PA and RW in a particular way.

	{	
%
%
   	\begin{table}[!t]  
	\centering  
	\fontsize{7.5}{8}\selectfont  
	\begin{threeparttable}  
		\caption{Investigation of GRL, LRL, PA and RW on CODE-Net. Tested on $96$ images from Rain1200.} 
		
		\label{tab:ablation_study}  			
		\begin{tabular}{ccc|p{0.5cm}<{\centering}p{0.5cm}<{\centering}p{0.5cm}<{\centering}p{0.5cm}<{\centering}|ccc}
			\toprule
			       &&&GRL&LRL  &PA &RW&&PSNR&\cr
			 \midrule
			&\nth{1}&&\False&\True  &\True &\True&&30.56&\cr  
			&\nth{2}&&\True&\False  &\False &\False&&29.40\cr  
			&\nth{3}&&\True&\False  &\False &\True&&30.00\cr  
			&\nth{4}&&\True&\False  &\True &\False&&29.34\cr  
			&\nth{5}&&\True&\False  &\True &\True&&29.92\cr  
			&\nth{6}&&\True&\True  &\False &\False&&29.85\cr  
			&\nth{7}&&\True&\True  &\False &\True&&30.88\cr  
			&\nth{8}&&\True&\True  &\True &\False&&29.58\cr  
			&\nth{9}&&\True&\True  &\True &\True&&\bf 31.05\cr  

			\bottomrule  
		\end{tabular} 
		
    	\end{threeparttable}  
	\end{table} 

%
%
}

	
	
	\subsection{Failure Cases}
	{Though the proposed network could achieve superior performance on most of testing images while producing corresponding RDEs, some situations occurred in which our method would produce overestimation of RDE and over deraining result, as exhibited in Fig.~\ref{fig:failure}. This is primarily caused by the confusing of rain-like structures and the crude global average pooling operation in LWB \myref{def:lwb} that only considers the overall information of each channle. We believe that the predicament could be tackled if we capture sufficient local details additionally.}
	
	\subsection{Extensions}
	In this section, we explore the potentials of our network for other visual tasks.  In the pursuit of better deraining effect, we first train a generative adversarial CODE-Net (GaCODE-Net) using perceptual loss~\cite{per-loss1,per-loss2}. As shown in Fig.~\ref{fig:extesions-visual-pursuit}, although inferior to CODE-Net on quantitative metrics (PSNR and SSIM), GaCODE-Net is capable of generating more realistic textures and high-frequency details. {Moreover, since snowy image could be modeled as \myref{eq:1} \cite{tmm2018tian} and snow streaks share much similarity with rain streaks, we then evaluate our CODE-Net on image desnowing task. Fig.~\ref{fig:extesions-desnowing} shows the results on two real snowy images, which verifies the promising generalization of our method on other tasks with similar degeneration model.}

	\section{Conclusion}
	\label{sec:Conclusion}
	
	{Single image deraining with unknown rain densities and multi-scale raindrops are very challenging and rarely considered in literatures. In this paper, a density guided network (CODE-Net) and its multi-scale version mCODE-Net are proposed, where both rain density and raindrop scale are considered. Representing rain streaks with the CSC model, we find that the sparse coefficients of rain streaks and the rain density are closely related. Consequently, the rain density is implicitly considered by learning weights from sparse coefficients through the channel attention blocks. And extending CODE-Net to multi-scale dictionaries, the challenge of multi-scale raindrops for SIDR can be straightforwardly addressed. Both qualitative and quantitative evaluations demonstrate  the superiority to recent state-of-the-art methods. Extensions to applications of other low-level vision tasks are also explored and show the generality of our proposed CODE-Net. 
		
	{\color{black}Moreover}, with the learned weights, we proposed a simple approach, namely RDE, to estimate the continuously valued rain density. The effectiveness of RDE is validated in various clean, rainy and derained images, while the potential for evaluating deraining algorithms is reserved for future works. Besides, to obtain more precise weights, we plan to improve the learning weight block , such as leveraging more local details, \textit{e.g.}, spatial structure and nonlocal self-similarity.}


	\ifCLASSOPTIONcaptionsoff
	\newpage
	\fi

	
	
	\bibliographystyle{IEEEtran}
	\bibliography{mybibfile}
	%

	

\end{document}